\documentclass[3p,times]{elsarticle}

\usepackage{ecrc}

\volume{00}

\firstpage{1}

\journalname{Nuclear Physics A}

\runauth{Nabi and B\"{o}y\"{u}kata}

\jid{procs}

\jnltitlelogo{Nuclear Physics A}

\CopyrightLine{2013}{Published by Elsevier Ltd.}

\usepackage{amssymb}
\usepackage{amsthm}
\usepackage{latexsym}
\usepackage{graphics}
\usepackage{graphicx}
\usepackage{epsfig}

\usepackage{rotating}

\begin{document}

\begin{frontmatter}



\dochead{}


\title{$\beta$-decay half-lives and nuclear structure of exotic proton-rich waiting point nuclei under $rp$-process conditions}


\author{Jameel-Un Nabi$^1$ and Mahmut B\"{o}y\"{u}kata$^2$}

\address{
$^1$Faculty of Engineering Sciences, GIK Institute of
Engineering Sciences and Technology, Topi 23640, Swabi, Khyber
Pakhtunkhwa, Pakistan}

\address{
$^2$Department of Elementary Science Education, Faculty of Education, \d{C}anakkele Onsekiz Mart University,  TR-17100, \c{C}anakkele, Turkey}

\begin{abstract}
We investigate even-even nuclei in the $A\sim70$ mass region within
the framework of the proton-neutron quasi-particle random phase
approximation (\mbox{pn-QRPA}) and the interacting boson model-1
(\mbox{IBM-1}). Our work includes calculation of the energy spectra
and the potential energy surfaces $V(\beta,\gamma)$ of Zn, Ge, Se,
Kr and Sr nuclei with the same proton and neutron number, $N = Z$.
The parametrization of the \mbox{IBM-1} Hamiltonian was performed
for the calculation of the energy levels in the ground state bands.
Geometric shape of the nuclei was predicted by plotting the
potential energy surfaces $V(\beta,\gamma)$ obtained from the
\mbox{IBM-1} Hamiltonian in the classical limit. The \mbox{pn-QRPA}
model was later used to compute half-lives of the neutron-deficient
nuclei which were found to be in very good agreement with the
measured ones. The \mbox{pn-QRPA} model was also used to calculate
the Gamow-Teller strength distributions and was found to be in
decent agreement with the measured data. We further calculate the
electron capture and positron decay rates for these $N$ = $Z$
waiting point (WP) nuclei in the stellar environment employing the
\mbox{pn-QRPA} model. For the $rp$-process conditions, our total
weak rates are within a factor two compared with the Skyrme
HF+BCS+QRPA calculation. All calculated electron capture rates are
comparable to the competing positron decay rates under $rp$-process
conditions. Our study confirms the finding that electron capture
rates form an integral part of the weak rates under $rp$-process
conditions and should not be neglected in the nuclear network
calculations.

\end{abstract}


\end{frontmatter}





\section{Introduction}
\label{sec:level1}

Modeling and simulation of the explosive phenomena involve the
knowledge of the properties of the exotic nuclei. X-ray bursts (see
e.g. \cite{Sch98, Wal81}) are generated by a thermonuclear runaway
in an hydrogen-rich environment where an accreting neutron star is
fed from a binary partner (usually a red giant). The ignition starts
as soon as the temperature $T$ and the density $\rho$ in the
accreted disk due to mass transfer become sufficiently high to
permit a breakout from the hot CNO cycle. Stellar temperatures in
the vicinity of $T$ = 1 -- 3 GK and densities in the range $\rho$ =
10$^{6}$ -- 10$^{7}$ gm.cm$^{-3}$ are believed to facilitate the
development of the nucleosynthsis along the proton-rich side via
rapid proton capture ($rp$) process (see e.g. \cite{Sch98, Pri00}).
Studies (e.g. \cite{Wor94}) have pointed toward extreme hydrogen
burning where at sufficiently high temperature (T$_{9} \leq 0.8 $K)
and density conditions ($\rho \geq 10^{4}$ gcm$^{-3}$), depending on
the time scale of the explosive event, the $rp$- and the
$\alpha$p-processes reaction path may well proceed beyond mass A =64
and Z = 32. The proton capture processes are orders of the magnitude
faster than $\beta$-decays or any other competing processes. Schatz
and collaborators \cite{Sch01} found out that the $rp$-process is
responsible for producing heavy proton-rich nuclei reaching mass
number 100 that ends in a closed SnSbTe cycle. It also explained the
luminosity and the energy profiles present in the X-ray bursts.

The reaction path for $rp$-process follows a series of rapid proton
capture reactions producing heavy nuclei. The process culminates
when a radioactive element is produced and later proceeds through a
$\beta$-decay (double proton capture are less probable). The
reaction flow then has to wait for a relatively slower $\beta$-decay
and this nucleus is termed as a waiting point (WP) nucleus. Typical
time scale of the $rp$-process is $\sim$ 100 $s$. Half-lives of the
WP nuclei is roughly of the same order and hence determine the time
scale of the nucleosynthesis process and isotopic abundances.
Consequently, weak-interaction mediated stellar rates of the
neutron-deficient medium-mass nuclei play a crucial role in a better
understanding of the $rp$-process. Authors in Ref. \cite{Sch98}
claimed that important parameters for a successful $rp$-process
nucleosynthesis calculation include nuclear masses, nuclear
deformations (specially in the regime A = 60 -- 80 because of the
wide variety of the nuclear shapes displayed in the region) and
finally a reliable calculation of the stellar electron capture and
$\beta$-decay rates of the WP nuclei along a given reaction path
since they determine time structure and abundance patterns.

New generation radioactive ion-beam facilities (e.g. FAIR (Germany),
FRIB (USA) and FRIB (Japan)) are expected to reveal some exciting
facts about exotic nuclei but the current status is that majority of
them are not well explored. Consequently, astrophysical simulations
of the violent and explosive events must be built on nuclear-model
predictions which in turn need to be microscopic and reliable. We
attempt first to calculate nuclear shape of the WP nuclei using a
microscopic model. In this article, the calculation of the
interacting boson model
(IBM)~\cite{Arima75,Arima76,Arima78a,Arima79} is performed for the
low-lying collective levels and the prediction of geometric shape of
the $N$~=~$Z$ WP nuclei. The investigation of the nuclear properties
of nuclei in various region of the nuclear chart within IBM model is
still one of the hot subjects. Many experimental data of nuclei in
the isotopic (or isotonic) chain have been compared with
the~\mbox{IBM} calculations (for recent works, see Refs.
~\cite{Gladnishki12,Albers13,Duckwitz13,Thomas13}). The prediction
of the geometric type of nuclei with the corresponding potential
energy surface (PES) of the IBM model is also favorite subject as
seen recently work in Refs. ~\cite{Boyukata10, Nomura11a, Nomura11b,
Nomura11c, Ramos14a, Ramos14b, Kotila14b}.

For the $A\sim70$ mass region, the structure of the energy spectra
of $^{64}$Ge isotope with neighbors up to $^{78}$Ge was investigated
within the extended IBM model~\cite{Hsi92} and a transition from the
mixture of $SU(3)$, $O(6)$, $U(5)$ symmetries (mixture of prolate,
$\gamma$-unstable, spherical shapes) to $O(6)$ and $U(5)$ (mixture
of $\gamma$-unstable, spherical shapes) and then finally $U(S)$
symmetry (spherical shape) appear along to isotopic chain of
even-even $^{64-78}$Ge. The positive-parity bands of the even-even
$^{72-84}$Kr isotopes were studied within the framework of the IBM-2
model~\cite{Gia00} by taking in to account the $U(5)$ symmetry of
the model as a starting point of the investigation.

For the potential energy surfaces (PESs), Hartree-Fock-Bogoliubov
(HFB) model have been applied to this region as follows: the PESs of
the $^{64}$Ge, $^{68}$Se, $^{72}$Kr, $^{76}$Sr, $^{80}$Zr and
$^{84}$Mo were calculated within the HFB model using the Skyrme
interaction ~\cite{Yam01} and in this work the shape transition
changes from triaxial for $^{64}$Ge, oblate for $^{68}$Se,
$^{72}$Kr, large prolate for $^{76}$Sr, $^{80}$Zr, to spherical for
$^{84}$Mo. The PESs of $^{64-72}$Ge were calculated with the HFB
model plus the GognyD1S interaction ~\cite{Cor13} and the results
indicate that $^{64}$Ge is rigid triaxial, $^{66}$Ge is
$\gamma$-unstable, $^{68-72}$Ge isotopes are soft triaxial. The
solutions of the HFB model were applied to $^{68}$Se and $^{72}$Kr
nuclei ~\cite{Kob05} to describe the oblate-prolate shape
coexistence phenomena in these nuclei by using the
pairing-plus-quadrupole (P+Q) interaction. The PESs of the
$^{72-76}$Kr were performed within the HFB method by using the D1S
interaction and the minima for $^{74,76}$Kr are located at axial
deformed while the absolute minimum is found at oblate deformation
for $^{72}$Kr~\cite{Gir09}. The oblate-prolate shape coexistence in
proton-rich $^{68,70,72}$Se is investigated within the constrained
HFB plus the local QRPA (CHFB+LQRPA) equation by using the P+Q
interaction~\cite{Hin10}. The PESs of the even-even Ni, Zn, Ge, Se,
Kr, Sr, Zr, Mo, Ru, Pd, Cd, and Sn nuclei around the $N=Z$ line
cover a whole proton shell ranging from $Z=28$ up to $Z=50$ have
been worked within the HF + BCS calculations plus the Skyrme force
SLy4 ~\cite{Sar11}. Both the lighter and heavier nuclei close to
$Z=28$ and $Z=50$ tend to be spherical. The deformed shapes appear
around the mid shell nuclei in this work.

The main purpose of this work is to investigate some nuclear structure
properties like energy levels, B(E2) values, PES and to calculate half-
lives and stellar weak rates of the WP nuclei. The energy levels in the
ground state band of the nuclei in $N$~=~$Z$ line were calculated to fit
essential IBM Hamiltonian parameters. Later the potential energy surfaces
of each nuclei were plotted to get the deformation parameter $\beta$.
The nuclear model chosen to calculate terrestrial and
stellar weak decay rates is the proton-neutron quasiparticle random
phase approximation (\mbox{pn-QRPA}) which has a proven track-record
for the calculation of electron capture and $\beta$-decay rates.
Half-lives of the $\beta^{-}$ decays were calculated systematically
for about 6000 neutron-rich nuclei between the beta stability line
and the neutron drip line using the \mbox{pn-QRPA} model
\cite{Sta90}. Similarly half-lives for $\beta^{+}$/electron capture
decays for neutron-deficient nuclei with atomic numbers Z = 10 - 108
were calculated up to the proton drip line for more than 2000 nuclei
using the same model \cite{Hir93}. These microscopic calculations
gave a remarkably good agreement with the then existing experimental
data (within a factor of two for more than 90$\%$ (73$\%$) of nuclei
with experimental half-lives shorter than 1 s for $\beta^{-}$
($\beta^{+}$/EC) decays). Later Nabi and Klapdor-Kleingrothaus
reported the calculation of the weak-interaction rates for more than
700 nuclei with A = 18 to 100 in stellar environment using the same
nuclear model \cite{Nab99}.

The paper is designed as follows: The IBM model is introduced and
its Hamiltonian with formalism of the potential energy surface are
presented in Section~2. The \mbox{pn-QRPA} model and formalism for
calculation of stellar rates are discussed in Section~3. We discuss
our calculation and compare with measured data and previous
calculations in Section~4. Conclusions are finally stated in
Section~5.

\section{The interacting boson model (IBM)}
\label{sec:level2}

The interacting boson model (IBM) of Arima and
Iachello~\cite{Arima75,Arima76,Arima78a,Arima79}, originally
applicable to even-even nuclei, is quite successful to describe the
collective properties of medium mass and heavy nuclei. The necessary
components of the IBM are the $s$ and the $d$ bosons with angular
momenta zero and two, respectively. The IBM model is also connected
to the nuclear shell model ~\cite{Talmi93} by the realization that
the $s$ and $d$ bosons can be interpreted as correlated Cooper pairs
formed by two valence nucleons in the shell coupled to angular
momenta $L=0$ and $L=2$. The simplest version of the model, making
no distinction between proton and neutron bosons, is called
as~\mbox{sd-IBM} as well as~\mbox{IBM-1} model. The extended version
of model that separates the proton and neutron bosons, is normally
referred to as proton-neutron~\mbox{(pn)-IBM} or~\mbox{IBM-2}
~\cite{Arima77,Otsuka78,Arima78b}. The algebraic~\mbox{IBM-1} model
is defined by a six-dimensional space because of the $s$ ($L=0$,
$\mu=0$) and the $d$ ($L=2$, $-2\leq$ $\mu$ $\leq2$) bosons. The
model needs a description of the $U(6)$ group structure. This group
presents three possible dynamical symmetries, labeled by $U(5)$,
$SU(3)$ and $O(6)$ limits, appearing in the following group chains:

    $U(6)\supset$  $U(5)\supset$ $O(5)\supset$ $O(3)$

    $U(6)\supset$  $SU(3)\supset$ $O(3)$

    $U(6)\supset$  $O(6)\supset$ $O(5)\supset$ $O(3)$. \\
Each of the three standard symmetries are related to the geometrical
shape of the nuclei; (i) the $U(5)$ limit corresponds to the
spherical, (ii) $SU(3)$ limit describes the axially deformed nuclei
and (iii) $O(6)$ limit defines $\gamma$-unstable (asymmetric
deformed) nuclei. One of the characteristic clue for each symmetries
is the energy ratio $R_{4/2}=E(4^{+})/E(2^{+})$ in the ground state
band. This ratio is $2.0$ for the spherical, $2.5$ for
$\gamma$-unstable and $3.33$ for the axially deformed nuclei.

\subsection{The IBM-1 Hamiltonian}
\label{sec:level3}

The \mbox{IBM-1} model describes a system of the $s$ and the $d$
bosons interactions, and therefore, the general \mbox{IBM-1}
Hamiltonian includes combinations of the operators $s$, $s^{\dag}$,
$d$, $d^{\dag}$~\cite{Arima76,Arima78a,Arima79}. This Hamiltonian
has six interactions in total, \emph{1} single-particle energy and
\emph{5} two-body interactions between the bosons. These
interactions are used as free parameters in the Hamiltonian and
these parameters must be fitted to the experimental data. The most
general Hamiltonian can also be written as the sum of the quadratic
Casimir operators of the subgroups of the complete $U(6)$ group
chain. However, the most commonly used type of the \mbox{IBM-1}
Hamiltonian, also used in this work, is called multipole expansion
in terms of the six parameters given in the following form ~\cite{Casten88};
\begin{equation}
\hat H= \epsilon\,\hat n_d+ a_0\,\hat P_+\cdot\hat P_- + a_1\,\hat
L\cdot\hat L+ a_2\,\hat Q\cdot\hat Q+ a_3\,\hat
T_3\cdot\hat T_3 + a_4\,\hat T_4\cdot\hat T_4, \label{e_ham}
\end{equation}
where $\hat n_d$ is the boson-number operator, $\hat P_+$ is the
boson pairing operator, $\hat L$ is angular momentum operator, $\hat
Q$ is quadrupole operator, and $\hat T_3$, $\hat T_4$ are the
octupole, hexadecapole operators, respectively and defined as
\begin{eqnarray}
\hat n_d&=&\sqrt{5}[d^\dag\times\tilde d]^{(0)}_0,
\nonumber\\
\hat P_+&=&[s^\dag\times s^\dag+\sqrt{5}\,d^\dag\times
d^\dag]^{(0)}_0, \qquad \hat P_-=\left(\hat P_+\right)^\dag,
\nonumber\\
\hat L&=&\sqrt{10}[d^\dag\times\tilde d]^{(1)},
\nonumber\\
\hat Q&=&[d^\dag\times\tilde s+s^\dag\times\tilde d]^{(2)}+
\overline{\chi}[d^\dag\times\tilde d]^{(2)},
\nonumber\\
\hat T_\ell&=&[d^\dag\times\tilde d]^{(\ell)}, \qquad \qquad \qquad
\ell=3, 4. \label{e_term}
\end{eqnarray}

Eq.~(\ref{e_ham}) has the advantage that the constants have been
used to describe the properties of a single nucleus by fitting to
experimental nuclear spectra~\cite{Iachello87}. The combination of
the linear and the quadratic operators of the $U(6)$ group and its
subgroups; $U(5)$, $SU(3)$, $O(6)$, $O(5)$, $O(3)$ can be written in
terms of the operators of the multipole form of the IBM-1
Hamiltonian given in Eqs. (\ref{e_ham}) and (\ref{e_term}). The
connections of the parameters for each forms of the Hamiltonian can
be reviewed in detail from Ref.~\cite{Casten88}. In addition to the
energy levels, the electric quadrupole transition rates can also be
calculated in the IBM-1 model by using the quadrupole transition
operator, $\hat T({\rm E2})=e_{\rm b} \cdot \hat Q$. Here, $e_{\rm
b}$ is the boson effective charge, $\hat Q$ is the quadrupole
operator and has the same role as shown in Eq. (\ref{e_term}).

\subsection{The geometry: The potential energy surface}
\label{sec:level4}

The investigation of the geometric character of the nucleus is one
of the interesting subjects in nuclear physics. The IBM model has a
relationship with the geometric model of Bohr and Mottelson
\cite{Bohr98}. Within the IBM model, the geometric shape of the
nuclei can be also visualized by plotting the potential energy
surface in terms of the deformation parameters $\beta$ and $\gamma$.
This potential energy surface $V(\beta,\gamma)$ can be obtained from
the \mbox{IBM-1} Hamiltonian in the classical
limit~\cite{Dieperink80a,Dieperink80b,Ginocchio80a,Ginocchio80b} as


\begin{eqnarray}
V(\beta,\gamma)=\epsilon N \frac{\beta^2}{1+\beta^2} +a_1 6N\frac{\beta^2}{1+\beta^2} + a_2 N\left[\frac{5+(1+\overline{\chi}^2)\beta^2}{1+\beta^2} +(N-1)\frac{\left(\frac{2\overline{\chi}^2\beta^4}{7}+4\sqrt{\frac{2}{7}}\overline{\chi}\beta^3\cos(3\gamma)+4\beta^2\right)}{(1+\beta^2)^2}\right]
\nonumber\\
+a_0\frac{N(N-1)}{4} \frac{1-\beta^2}{1+\beta^2}
+a_3 7N \frac{(5+\overline{\chi}^2)}{5}\frac{\beta^2}{1+\beta^2}
+a_4 N \left[\frac{9}{5}\frac{\beta^2}{1+\beta^2}+\frac{18}{35} (N-1) \frac{\beta^4}{(1+\beta^2)^2}\right].
\label{e_pes}
\end{eqnarray}

This formula is reproduced for the most general \mbox{IBM-1}
Hamiltonian~(\ref{e_ham}). Both Eqs~(\ref{e_ham}) and (\ref{e_pes})
include common parameters, except for the deformation parameters.
Here, $N$ is number of the bosons and $\overline{\chi}$ belongs to the
quadrupole operator given in Eq.~(\ref{e_term}). The $\beta$ and
$\gamma$ variables play the same role as in the geometric collective
model (GCM). For spherical nuclei, $\beta=0$, $\gamma=0^\circ$ and for the
deformed nuclei; $\beta\neq0$ and $\gamma=0^\circ$, $30^\circ$,
$60^\circ$ for prolate, triaxial, and oblate shapes, respectively.
However, the classical limit for the Hamiltonian of each limit:
$U(5)$, $U(5)$, $U(5)$, can also be derived as shown in
Refs.~\cite{Dieperink80a,Isacker81}.

\section{The \mbox{pn-QRPA} formalism and stellar weak rates}
\label{sec:level5}

In our \mbox{pn-QRPA} model, single particle energies and wave
functions were calculated using the Nilsson model \cite{Nil55}, as
it takes into account nuclear deformation. Pairing correlation was
treated in the BCS approximation. The proton-neutron residual
interactions occur through two channels, namely as particle-hole and
particle-particle interactions. These interactions were given
separable form and were characterized by two interaction constants
$\chi$ and $\kappa$, respectively.  We wanted to incorporate nuclear
deformation parameter ($\beta$)  from the IBM-1 model. However
$\beta$ in IBM model plays a slightly different role than $\beta$
does in different geometric models. Whereas in the IBM $\beta$
describes the quadrupole mixing of only the $2N$ valence nucleons
(the remaining $A-2N$ are in a spherical core), $\beta$ in
geometrical models refers to the deformation of all $A$ nucleons
\cite{Ginocchio80b}. Then the bosons in the IBM approximate fermion
pairs. So naturally the properties of the nucleus will depend on the
structure of these pairs. The deformations in the two models are
connected by $\beta_{geom}\lesssim 1.18(2N/A)\beta_{IBM}$ as
explained in detail in Ref. ~\cite{Ginocchio80b}.  Nomura and
collaborators \cite{Nom10} studied this relationship and tried to
relate the IBM parameters with the microscopic selfconsistent
mean-field calculations. They found that the $\beta$ parameters in
the two models were related roughly within a factor 3--5
\cite{Nom10}. It was apparent that linking the two deformation is
not so simple. The deformation parameter was argued to be one of the
most important parameters in pn-QRPA calculations \cite{Sta90,Ste04}
and therefore we decided to use the $\beta$ parameter from the
relativistic mean-field calculation by Lalazissis and collaborators
\cite{Lal99}. Calculated $\beta$ parameters within the IBM-1 model
and relativistic mean-field model are shown in Table~\ref{ta1ii}.
The particle-hole and particle-particle interactions strength
parameters, $\chi$ and $\kappa$, are regarded as the two most
important model parameters in the \mbox{pn-QRPA} theory (see Refs.
\cite{Sta90,Hir93}). In this work, $\kappa$ was fixed at 0.1 MeV.
The particle-hole interaction parameter $\chi$ is known to affect
the position of the Gamow-Teller giant resonance and was set using a
$1/A$ dependence \cite{Hir93}. We were able to deduce a value of
$4.2/A$ as the optimum value of $\chi$  which best reproduced the
experimental half-lives \cite{Aud12} using the deformation parameter
from \cite{Lal99}. It is to be noted that a previous pn-QRPA
calculation \cite{Nab12} used different values for $\chi$ and
$\kappa$ for the case of $^{72}$Kr and $^{76}$Sr because in that
calculation deformation parameter was used from \cite{Moe81}.  For
further discussion on choice of these two parameters we refer to
\cite{Hir93} and references therein. Nuclear masses and Q-values
required for the calculation were taken from the recent atomic mass
evaluation AME2012 \cite{Aud12}.

The electron capture (ec) and the positron decay (pd) rates of a
transition from the $i^{th}$ state of the parent to the $j^{th}$
state of the daughter nucleus are given by
\begin{equation}
\lambda ^{^{ec(pd)} } _{ij} =\left[\frac{\ln 2}{D}
\right]\left[B(F)_{ij} +\left({\raise0.7ex\hbox{$ g_{A}
$}\!\mathord{\left/ {\vphantom {g_{A}  g_{V} }} \right.
\kern-\nulldelimiterspace}\!\lower0.7ex\hbox{$ g_{V}  $}}
\right)^{2} B(GT)_{ij} \right]\left[f_{ij}^{ec(pd)} (T,\rho ,E_{f}
)\right]. \label{phase space}
\end{equation}
The value of D was taken to be 6295s \cite{Yos88}. B(F) and B(GT)
are reduced transition probabilities of the Fermi and ~Gamow-Teller
(GT) transitions, respectively,
\begin{equation}
B(F)_{ij} = \frac{1}{2J_{i}+1} \mid<j \parallel \sum_{k}t_{\pm}^{k}
\parallel i> \mid ^{2}.
\end{equation}
\begin{equation}
B(GT)_{ij} = \frac{1}{2J_{i}+1} \mid <j \parallel
\sum_{k}t_{\pm}^{k}\vec{\sigma}^{k} \parallel i> \mid ^{2}.
\end{equation}
Here $\vec{\sigma}^{k}$ is the spin operator and $t_{\pm}^{k}$
stands for the isospin raising and lowering operator with
$(g_{A}/g_{V})$ = -1.254 \cite{Rod06}. Details of the calculation of
the reduced transition probabilities can be found in Ref.
\cite{Nab99a, Nab04}. We used a quenching factor of 0.6 in our
calculation as normally employed in most shell model and QRPA
calculations of the weak rates.

The $f_{ij}^{ec(pd)}$ in Eq.~(\ref{phase space}) are the phase space
integrals and are functions of the stellar temperature ($T$), the
density ($\rho$) and Fermi energy ($E_{f}$) of the electrons. They
are explicitly given by
\begin{equation}
f_{ij}^{ec} \, =\, \int _{w_{l} }^{\infty }w\sqrt{w^{2} -1}
 (w_{m} \, +\, w)^{2} F(+Z,w)G_{-} dw,
 \label{ec}
\end{equation}
and by
\begin{equation}
f_{ij}^{pd} \, =\, \int _{1 }^{w_{m}}w\sqrt{w^{2} -1} (w_{m} \,
 -\, w)^{2} F(- Z,w)(1- G_{+}) dw.
 \label{pd}
\end{equation}
In Eqs.~(\ref{ec}) and ~(\ref{pd}), $w$ is the total energy of the
electron including its rest mass. $w_{m}$ is the total $\beta$-decay
energy,
\begin{equation}
w_{m} = m_{p}-m_{d}+E_{i}-E_{j},
\end{equation}
where $m_{p}$ and $E_{i}$ are masses and excitation energies of the
parent nucleus, and $m_{d}$ and $E_{j}$ of the daughter nucleus,
respectively. F($ \pm$ Z,w) are the Fermi functions and were
calculated according to the procedure adopted by Gove and Martin
\cite{Gov71}. G$_{\pm}$ are the Fermi-Dirac distribution functions
for positrons (electrons).
\begin{equation}
G_{+} =\left[\exp \left(\frac{E+2+E_{f} }{kT}\right)+1\right]^{-1},
\end{equation}
\begin{equation}
 G_{-} =\left[\exp \left(\frac{E-E_{f} }{kT}
 \right)+1\right]^{-1},
\end{equation}
here $E$ is the kinetic energy of the electrons and $k$ is the
Boltzmann constant.

The total decay rate per unit time per nucleus was calculated using
\begin{equation}
\lambda^{ec(pd)} =\sum _{ij}P_{i} \lambda _{ij}^{ec(pd)},
\label{tot}
\end{equation}
where $P_{i}$ is the probability of the occupation of the parent
excited states and follows the normal Boltzmann distribution. After
the calculation of all partial rates for the transition $i
\rightarrow j$ the summation was carried out over 200 initial (up to
10 MeV in parent nucleus) and 300 final states (up to 30 MeV in
daughter nucleus) and satisfactory convergence was achieved in the
rate calculation. The convergence was attributed to a spacious model
space of 7$\hbar\omega$ in our calculation. The excited states of an
even-even nucleus in the current pn-QRPA model are two-proton and
two-neutron quasiparticle states. On the other hand the daughter
excited states are constructed as two-quasiparticle or
four-quasiparticle states. Collective states cannot be calculated in
the model and is a shortcoming of the current pn-QRPA model.  In
order to make up for this deficiency and to further increase the
reliability of the calculated rates, experimental data were
incorporated in the rate calculation wherever possible.  The
calculated excitation energies, using the \mbox{pn-QRPA} model, were
replaced with measured levels when they were within 0.5 MeV of each
other. Two missing 1$^{+}$ measured states in $^{60}$Cu at 0.062 MeV
and 0.346 MeV (along with their log$ft$ values of 5.33 and 5.93,
respectively) were inserted manualy. No theoretical levels were
replaced with the experimental ones beyond the excitation energy for
which experimental compilations had no definite spin and/or parity.

\section{Results and discussions}
\label{sec:level6}

In the present application to the $A\sim70$ region, one of the
simplest form of \mbox{IBM-1} Hamiltonian has been used and given as
\begin{equation}
\hat H= \epsilon\,\hat n_d+ a_2 \,\hat Q\cdot\hat Q.
\label{e_s.ham}
\end{equation}
Here, $\hat n_d$ and $\hat Q$ are the boson-number and quadrupole
operators, respectively, defined in Eq. (\ref{e_term}). It is to be
noted that we have three parameters, $\epsilon$, $a_2$, and
$\overline{\chi}$ (in the quadrupole operator given in Eq. (\ref{e_term})).

The parametrization of the Hamiltonian constants  was performed for
the nuclei $^{60}$Zn, $^{64}$Ge, $^{68}$Se, $^{72}$Kr and $^{76}$Sr
located along the $N$ = $Z$ line in the $A\sim70$ mass region.
Parameters were firstly fitted for the lightest $^{60}$Zn nucleus
and then later expanded up to $^{76}$Sr, step by step.

\begin{figure}
\begin{center}
\includegraphics[width=12cm]{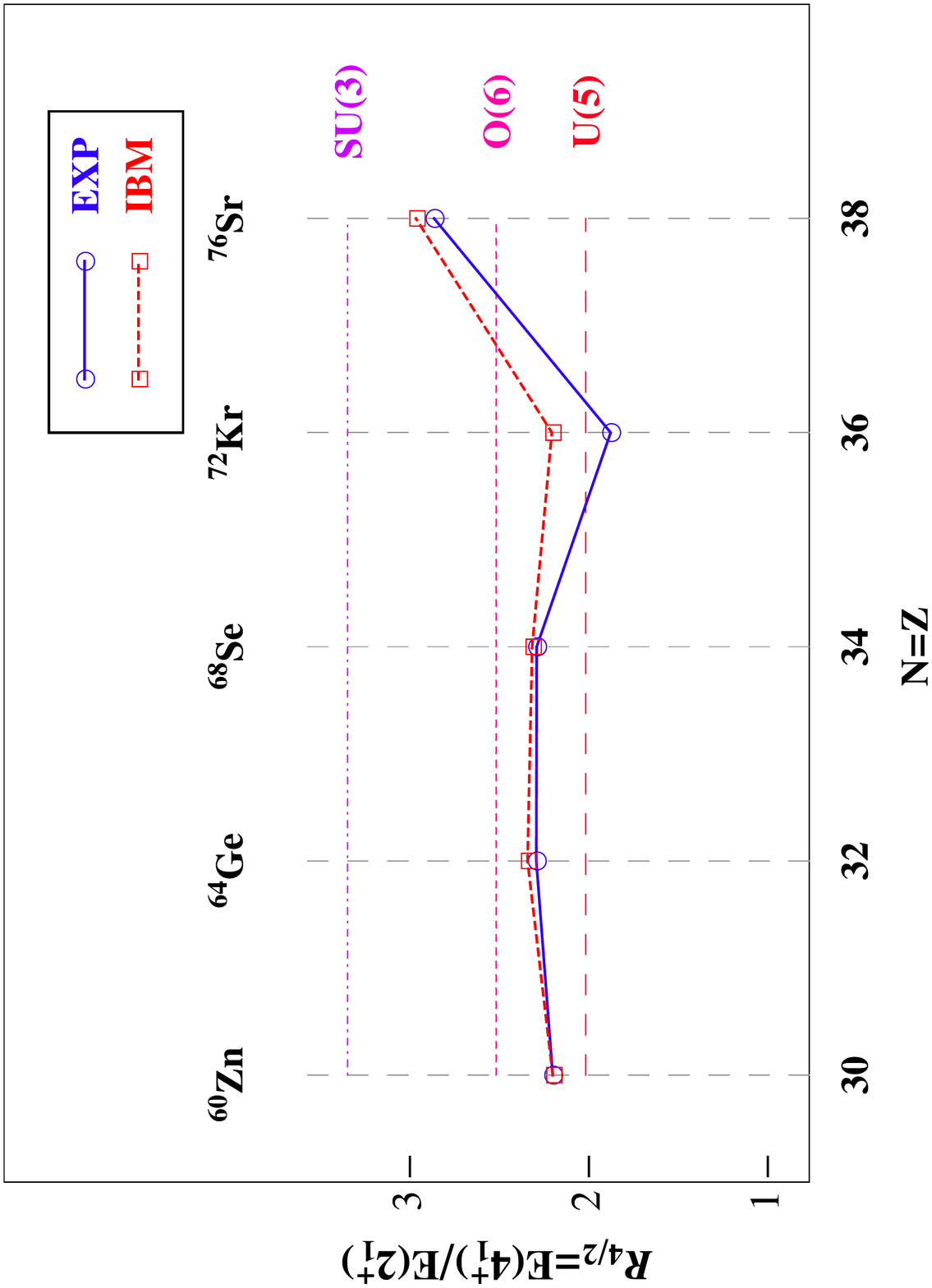}
\end{center}
\caption{(Color online) Values of the energy ratio
$R_{4/2}=E(4^+_1)/E(2^+_1)$ for $N$ = $Z$ nuclei.} \label{f_en2}
\end{figure}

\begin{figure}
\begin{center}
\includegraphics[width=15cm]{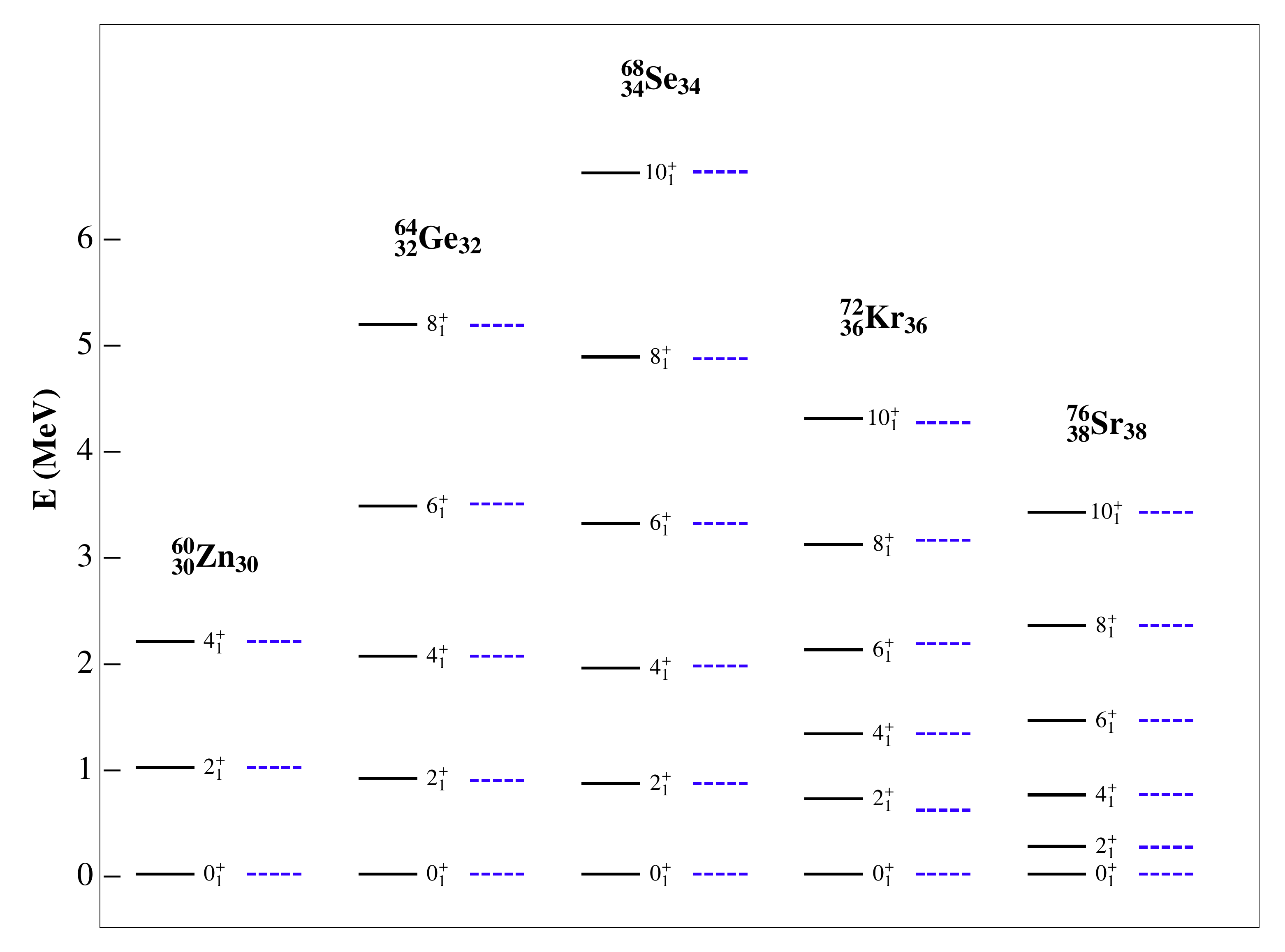}
\end{center}
\caption{(Color online) The experimental (straight lines) and
calculated (dashed lines) energy spectra for ground-state bands of
$^{60}$Zn, $^{64}$Ge, $^{68}$Se, $^{72}$Kr and $^{76}$Sr nuclei.}
\label{f_en1}
\end{figure}

The known energy levels in the ground state band of these nuclei
were selected to fitted constant parameters of the Hamiltonian by
minimizing the root-mean-square (\emph{rms}) deviation. First,
$\epsilon$ was adjusted and then $a_2$ by minimizing the \emph{rms}
deviation for each nucleus. Finally, same procedure was adopted for
the $\overline{\chi}$ value for each nucleus. The determined
parameters are given in Table~\ref{ta1}. It can be seen from
Table~\ref{ta1} that the $\overline{\chi}$ value of $^{72}$Kr
nucleus is higher than other $\overline{\chi}$ values. Detailed
fitted procedure for $\overline{\chi}$ value was re-performed by
keeping the same values for other two parameters ($\epsilon$ and
$a_2$) and, as shown in Table~\ref{ta1i}, minimum \emph{rms} was
obtained at $\overline{\chi}=-1.3$. The subsequent results were
sensible and shown in Fig.~\ref{f_en2}. In this figure, the energy
ratio $R_{4/2}=E(4^{+})/E(2^{+})$ in the ground state band is
illustrated for the studied nuclei along with typical values of the
$U(5)$, $SU(3)$ and $O(6)$ symmetries. The ratio of the $4^{+}$ and
$2^{+}$ yrast state energies indicates shape deformation in
even-even nuclei and a smaller value of this ratio implies a less
deformed nucleus \cite{Cas93}. It can be clearly seen from
Fig.~\ref{f_en2} that the $R_{4/2}$ values of $^{60}$Zn, $^{64}$Ge
and $^{68}$Se nuclei are almost same and located in between $U(5)$
and $O(6)$ symmetries and their fitted $\overline{\chi}$ values are
also same. However the measured $R_{4/2}$ value of $^{72}$Kr is
smaller than $2.0$ and positioned under the value of the $U(5)$ as
its $\overline{\chi}$ value is quite different. The $R_{4/2}$ value
of $^{76}$Sr lies in between $O(6)$ and $SU(3)$ limits. The overall
calculated results for low-lying energy spectra of given nuclei are
displayed in Fig.~\ref{f_en1}. Experimental data were taken from the
National Nuclear Data Center~\cite{NNDC}. Using the simple IBM-1
model, we may infer that $^{60}$Zn, $^{64}$Ge, $^{68}$Se and
$^{72}$Kr nuclei could be spherical since their energy ratios,
illustrated in Fig.~\ref{f_en2}, are around 2.0 and are close to
U(5). The $R_{4/2}$ value of $^{76}$Sr is close to $3.33$ and so
this nucleus is axially deformed.

The experimentally known B(E2: $2^+_1\rightarrow0^+_1$) values of
$^{68}$Se and $^{72}$Kr nuclei were also calculated using the boson
effective charge $e_{\rm b} = 0.097~eb$, fitted for $A\sim100$
region given in Ref.~\cite{Boyukata10}, and $\overline{\chi}$ values given in
Table \ref{ta1}. The calculated B(E2) values along with their
comparison with experimental data, taken from the National Nuclear
Data Center~\cite{NNDC}, are shown in Table \ref{ta1iii}.

As discussed above, the energy ratio is one of the important
signatures for the geometric behavior of a given nucleus. However,
other useful techniques include looking at the potential energy
surface based on the coherent state formalism. For this work, the
$V(\beta,\gamma)$ is reduced for the Hamiltonian as follows


\begin{eqnarray}
V(\beta,\gamma)=\epsilon \frac{N\beta^2}{1+\beta^2} + a_2 N\left[\frac{5+(1+\overline{\chi}^2)\beta^2}{1+\beta^2}
+(N-1) \frac{\left(\frac{2\overline{\chi}^2\beta^4}{7}+4\sqrt{\frac{2}{7}}\overline{\chi}\beta^3\cos(3\gamma)+4\beta^2\right)}{(1+\beta^2)^2}\right]
\label{e_s.pes}
\end{eqnarray}

\begin{figure}
\begin{center}
\includegraphics[width=15cm]{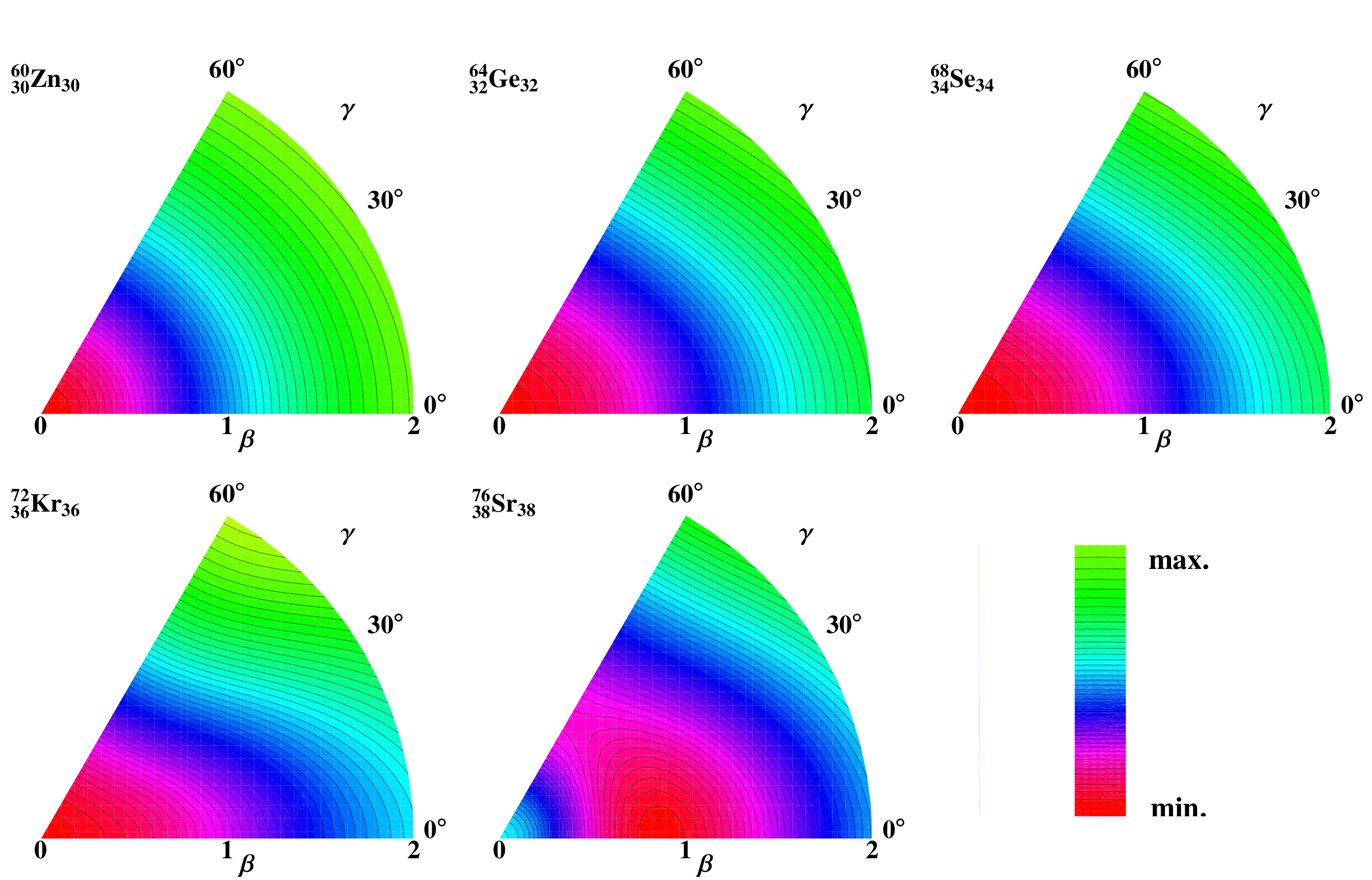}
\end{center}
\caption{(Color online) Potential energy surfaces for $^{60}$Zn,
$^{64}$Ge, $^{68}$Se, $^{72}$Kr and $^{76}$Sr nuclei.} \label{f_pes}
\end{figure}

The potential energy surfaces are illustrated in Fig.~\ref{f_pes}
for all nuclei included in the fit. As seen from the counterplot,
the shapes of  $^{60}$Zn, $^{64}$Ge, $^{68}$Se, and $^{72}$Kr are
spherical but the shape of $^{76}$Sr is deformed as also expected
from the signature of the energy ratio. As given in
Table~\ref{ta1ii}, $\beta_{IBM}=0.86$ for $^{76}$Sr. This value
appears rather bigger but this result is reasonable for the IBM
calculation since the SU(3) limit of $\beta_{min}$ approaches
$\sqrt{2}$ for large $N$ ~\cite{Ginocchio80a}.  According to the HFB
calculation by using different interactions (e.g. Skyrme
~\cite{Yam01,Sar11}, GognyD1S ~\cite{Cor13,Gir09},
pairing-plus-quadrupole (P+Q)~\cite{Kob05,Hin10}) for PESs in this
region, $^{64}$Ge is triaxial~\cite{Yam01} or rigid
triaxial~\cite{Cor13}, $^{68}$Se is oblate~\cite{Yam01}, $^{72}$Kr
is oblate~\cite{Yam01,Gir09} and $^{76}$Sr is prolate~\cite{Yam01}.

\begin{figure}
\begin{center}
\includegraphics[width=15cm]{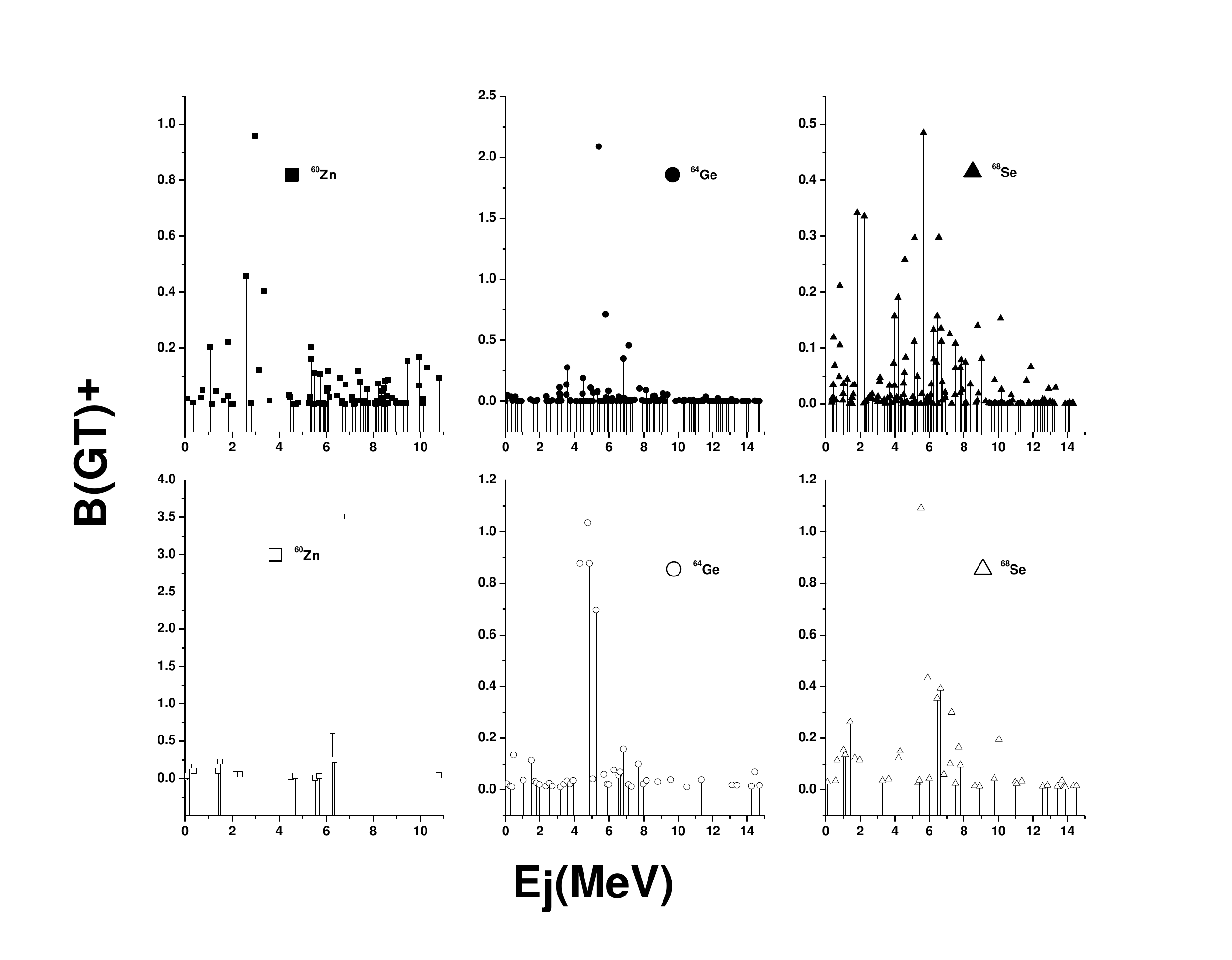}
\end{center}
\caption{Calculated BGT$^{+}$ strength distribution for $^{60}$Zn,
$^{64}$Ge and $^{68}$Se. Upper panels show calculation of this work
and lower panels those performed by Sarriguren \cite{Sar11}.}
\label{60-68}
\end{figure}

We compare our calculated BGT$^{+}$ strength distribution for
$^{60}$Zn, $^{64}$Ge and  $^{68}$Se against those calculated by
Sarriguren \cite{Sar11} in Fig.~\ref{60-68}. Details of the
formalism of Sarriguren's calculation may be seen in
\cite{Sar09a,Sar09b}. It is to be noted that Sarriguren used a
quenching factor of 0.55 in his calculation. As mentioned earlier we
incorporated a quenching factor of 0.6 in our calculation. Here the
upper panels depict our calculation whereas the lower panels display
the corresponding calculation by Sarriguren. The peak strengths in
the daughter 1$^{+}$ states match well  for the case of $^{64}$Ge
and $^{68}$Se. We calculated a total BGT$^{+}$ strength of 5.08 (up
to 10.8 MeV in daughter) for $^{60}$Zn, 6.17 (up to 14.7 MeV in
daughter) for $^{64}$Ge and 5.73 (up to 14.4 MeV in daughter) for
$^{68}$Se. The corresponding strengths calculated by Sarriguren are
5.27, 5.03 and 4.96, respectively.

\begin{figure}
\begin{center}
\includegraphics[width=15cm]{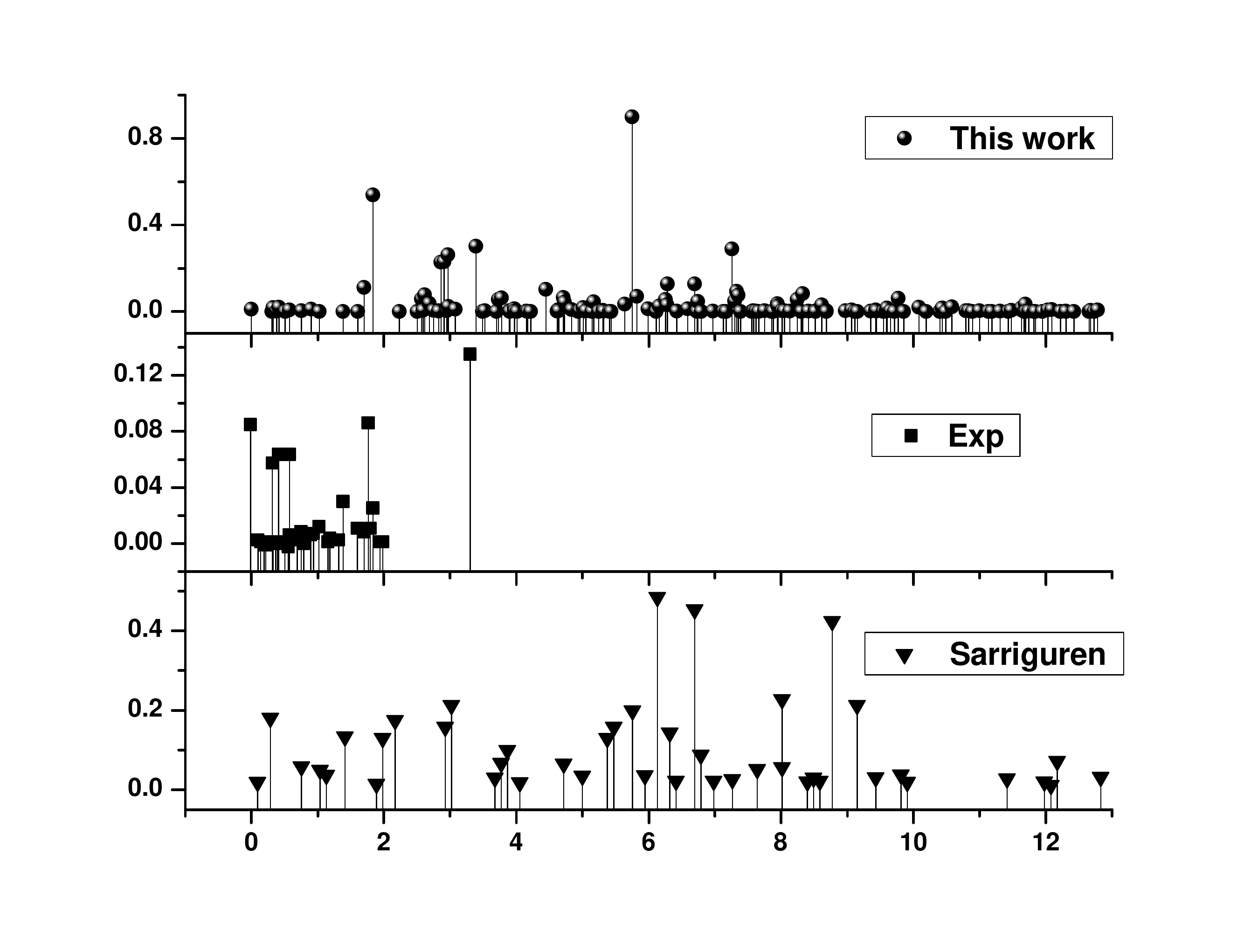}
\end{center}
\caption{BGT$^{+}$ strength distribution for $^{72}$Kr. Experimental
data taken from \cite{Piq03}. Lower panel shows calculation by
\cite{Sar11}.} \label{72}
\end{figure}

Experimental GT strength distribution for $^{72}$Kr was studied at
the ISOLDE PSB facility at CERN up to 3.3 MeV in $^{72}$Br
\cite{Piq03}. The measured GT strength distribution could only be
established up to 2 MeV as only one additional level was established
firmly at 3.3 MeV in $^{72}$Br. Fig.~\ref{72} shows our calculated
BGT$^{+}$ strength distribution for $^{72}$Kr (upper panel). The
middle panel shows the measured data and the bottom panel displays
the Sarriguren calculation. The total strength calculated by
Sarriguren up to 12.8 MeV in $^{72}$Br is 4.56. This is to be
compared with our calculated value of 4.96. Of special mention is
the peak that we calculate at 3.3 MeV. The measured data also shows
its peak at same excitation energy of 3.3 MeV in $^{72}$Br.

\begin{figure}
\begin{center}
\includegraphics[width=15cm]{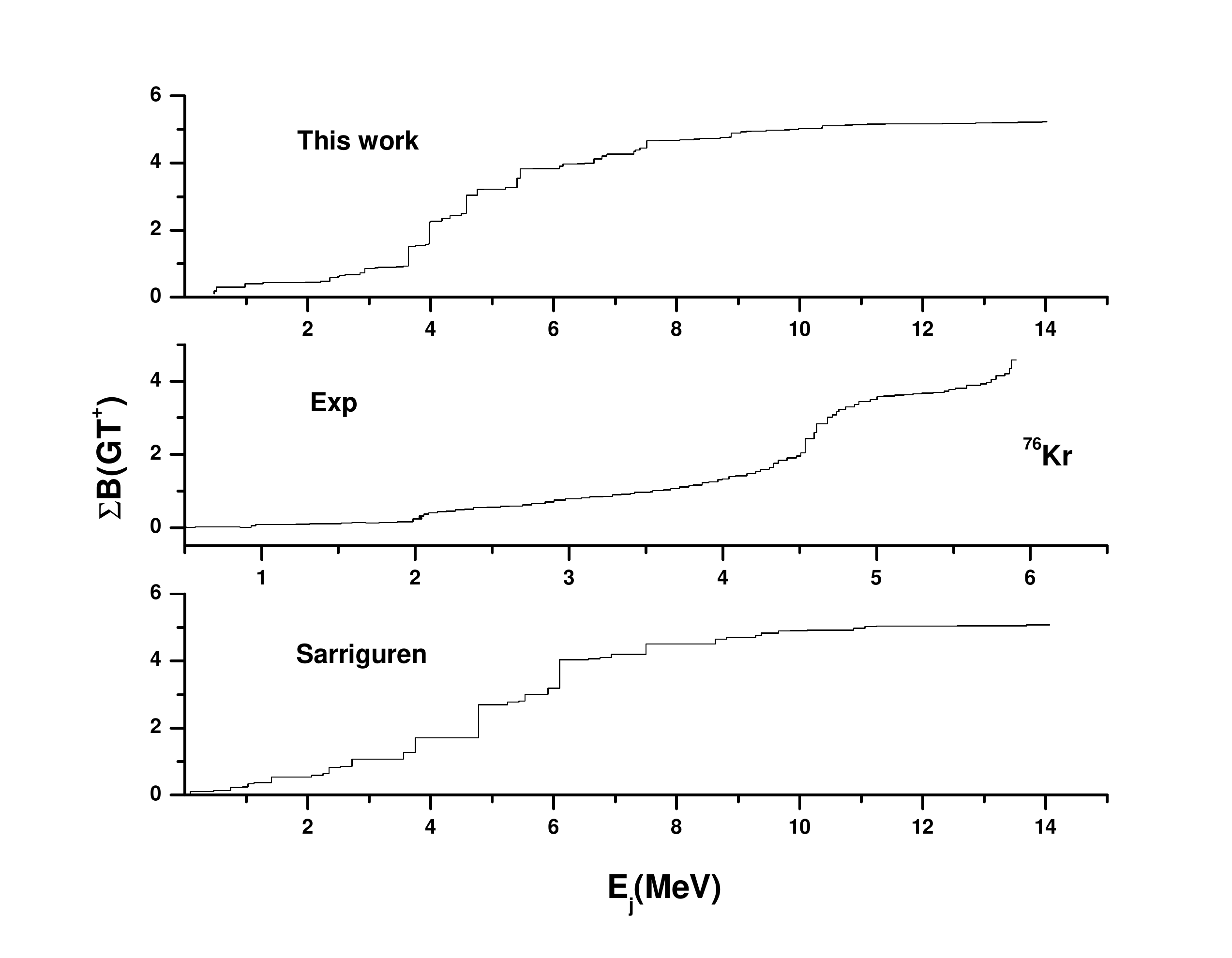}
\end{center}
\caption{Cumulative BGT$^{+}$ strength distribution for $^{76}$Sr.
Experimental data taken from \cite{Nac04}. Lower panel shows
calculation by \cite{Sar11}.} \label{76}
\end{figure}

The experimental GT strength distribution of the $N$ = $Z$ nucleus
$^{76}$Sr was also studied at CER-ISOLDE \cite{Nac04}. We compare
the experimental data with our calculation and those performed by
Sarriguren in Fig.~\ref{76}. Here we present the cumulative
BGT$^{+}$ strength distribution. Sarriguren calculated a total
strength of 5.07 up to 14.1 MeV to be compared with our value of
5.23. The experimental data bagged a total strength of 4.58.

\begin{figure}
\begin{center}
\includegraphics[width=4.3in,height=4.3in]{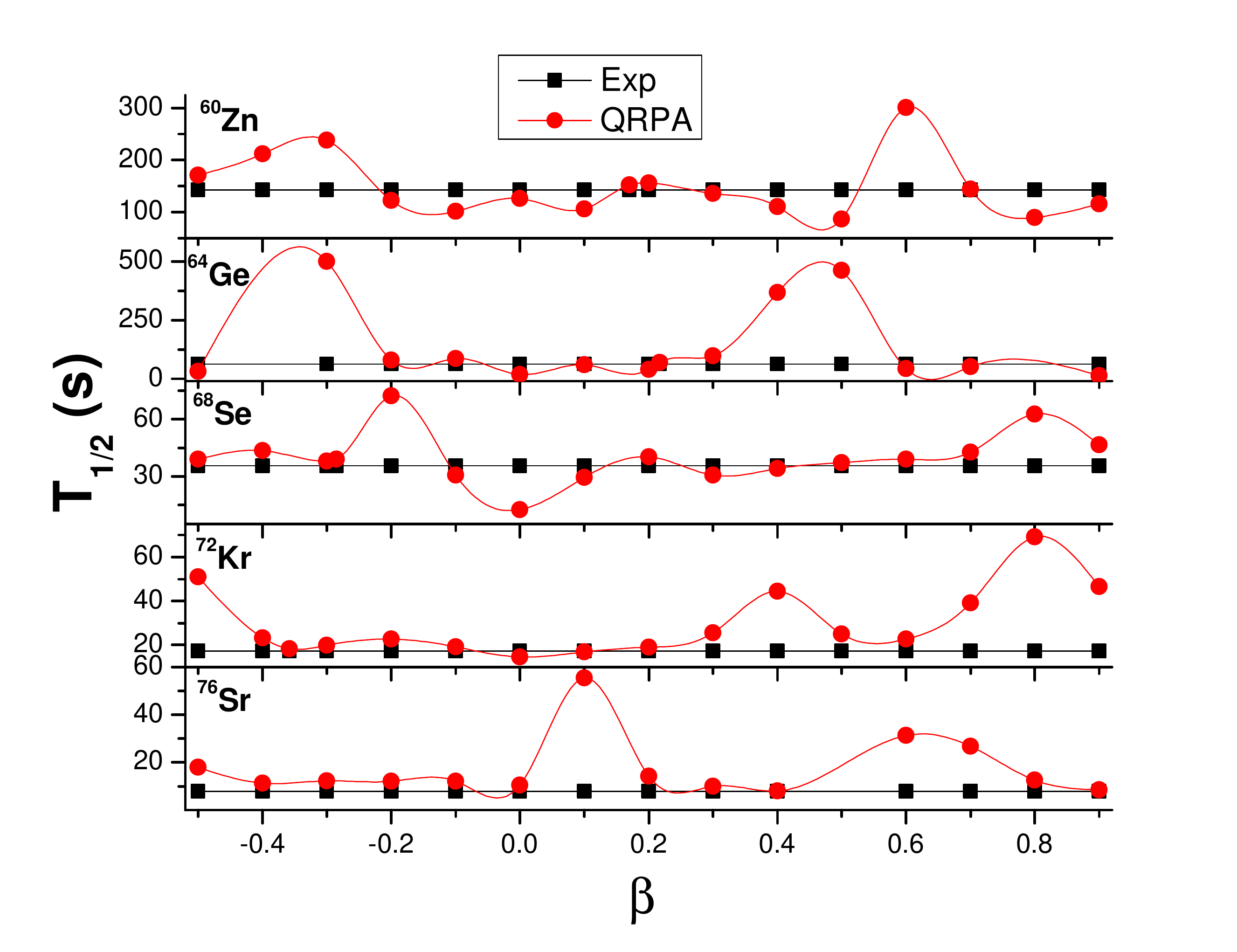}
\caption {Dependence of calculated half-lives on the values of
deformation parameter for $N$ = $Z$ nuclei.} \label{fig1}
\end{center}
\end{figure}

The variation of the calculated half-lives verses the deformation
parameter is shown in Fig.~\ref{fig1}. Experimental half-lives are
shown as black boxes whereas the calculated values are shown in
circles. We varied the value of the deformation parameter $\beta$
from $-0.5$ to $0.9$ in our calculation. It can be seen from
Fig.~\ref{fig1} that the calculated half-lives are sensitive
function of the nuclear deformation parameter. Further the
deformations from \cite{Lal99}  reproduce well the measured
half-lives of these $N = Z$ nuclei.

\begin{figure}
\begin{center}
\includegraphics[width=5.0in,height=4.3in]{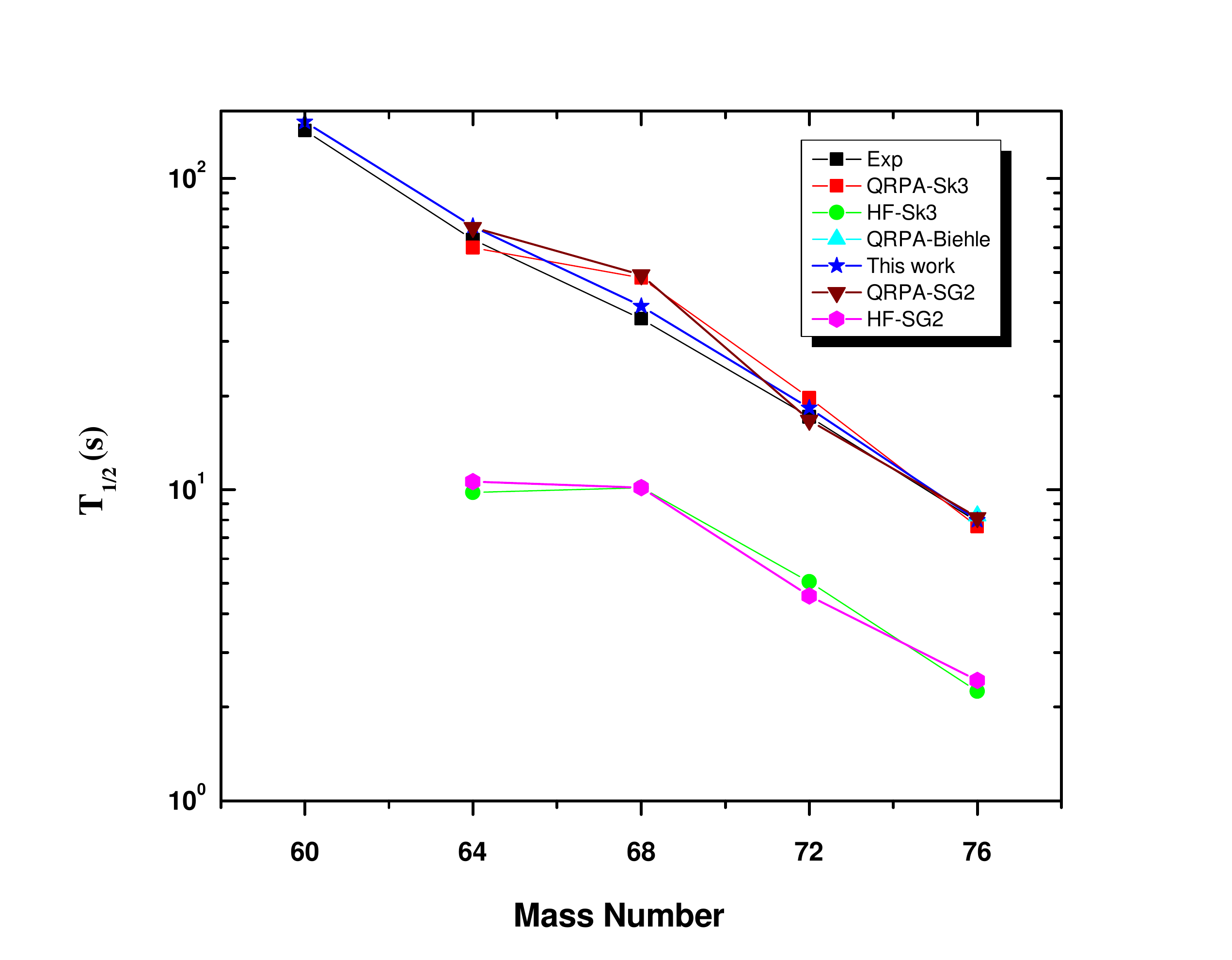}
\caption {(Color online) Comparison of experimental half-lives for
WP nuclei with this work and previous calculations.} \label{fig2}
\end{center}
\end{figure}

We compare our calculated half-lives for these neutron-deficient
medium-mass WP nuclei with experimental and previous calculations in
Fig.~\ref{fig2}. Experimental half-lives were taken from the recent
atomic mass evaluation AME2012 \cite{Aud12}. Shown also in
Fig.~\ref{fig2} are the Hartree-Fock (HF) and QRPA calculations
using the Sk3 \cite{Bei75} and SG2 \cite{Gia81} forces performed by
Sarriguren and collaborators \cite{Sar05}. Biehle and Vogel
\cite{Bie92} also performed a QRPA calculation for $^{76}$Sr, which
is in very good agreement with measured half-life, and is also shown
in Fig.~\ref{fig2}. It is noted that HF half-lives are
systematically lower than the corresponding QRPA and experimental
values. It is well known fact that the QRPA correlations tend to
reduce the mean-field Gamow-Teller strength thereby increasing the
calculated half-life values. We calculated a half-life of 152.3 $s$
for $^{60}$Zn to be compared with the experimental value of 142.8
$s$. Other models did not calculate half-life of $^{60}$Zn. Our
calculated half-lives are in good agreement with experimental
values. Our calculated percentage deviation from measured values for
$^{60}$Zn, $^{64}$Ge, $^{68}$Se, $^{72}$Kr and $^{76}$Sr are
6.7$\%$, 10.5$\%$, 9.7$\%$, 6.8$\%$ and 1.0$\%$, respectively.

\begin{figure}
\begin{center}
\includegraphics[width=4.3in,height=4.3in]{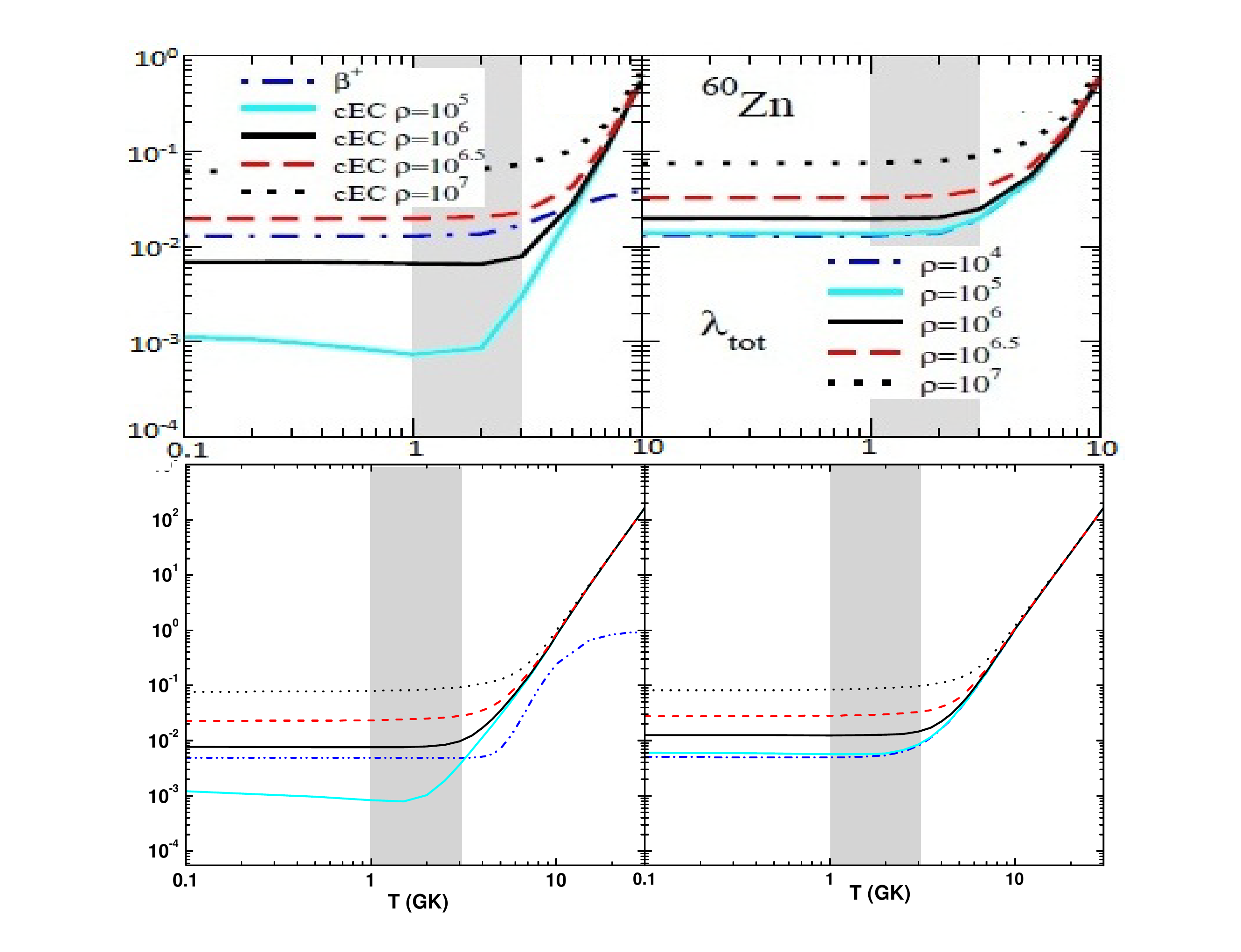}
\caption {(Color online) Comparison of calculated weak-interaction
mediated rates for $^{60}$Zn as a function of stellar temperature
and density. The upper panels show the deformed Skyrme Hartree-Fock
+ BCS  + QRPA calculation reproduced from \cite{Sar11}. The bottom
panels show the reported \mbox{pn-QRPA} calculation. The left panels
show the electron capture and positron decay rates whereas the right
panels show the combined total rates. All rates are given in units
of $s^{-1}$. Densities are given in units of $gcm^{-3}$ and
temperatures in units of $10^{9}$ K.} \label{fig3}
\end{center}
\end{figure}

\begin{figure}
\begin{center}
\includegraphics[width=4.3in,height=4.3in]{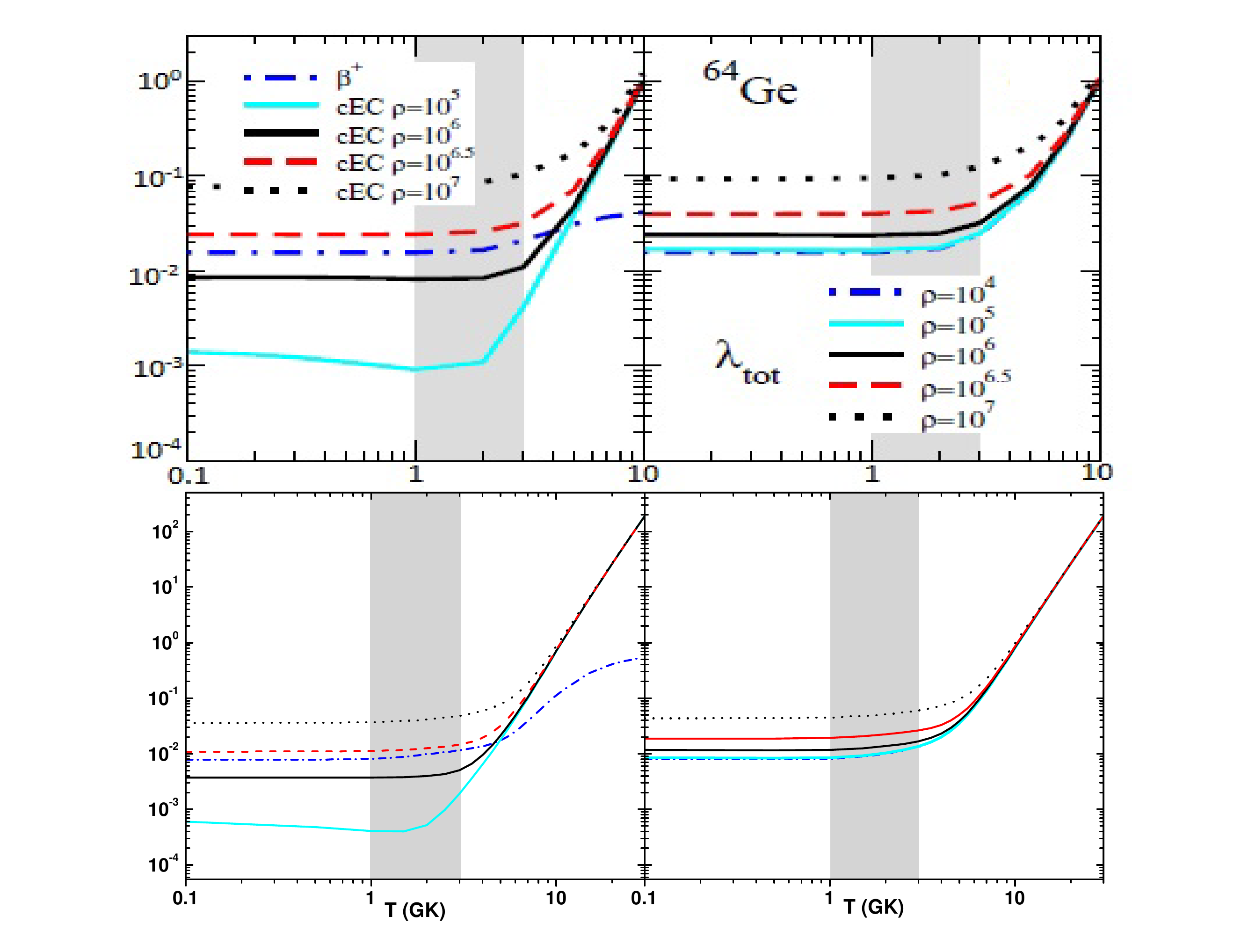}
\caption {(Color online) Same as Fig.~\ref{fig3} but for $^{64}$Ge.}
\label{fig4}
\end{center}
\end{figure}

\begin{figure}
\begin{center}
\includegraphics[width=4.3in,height=4.3in]{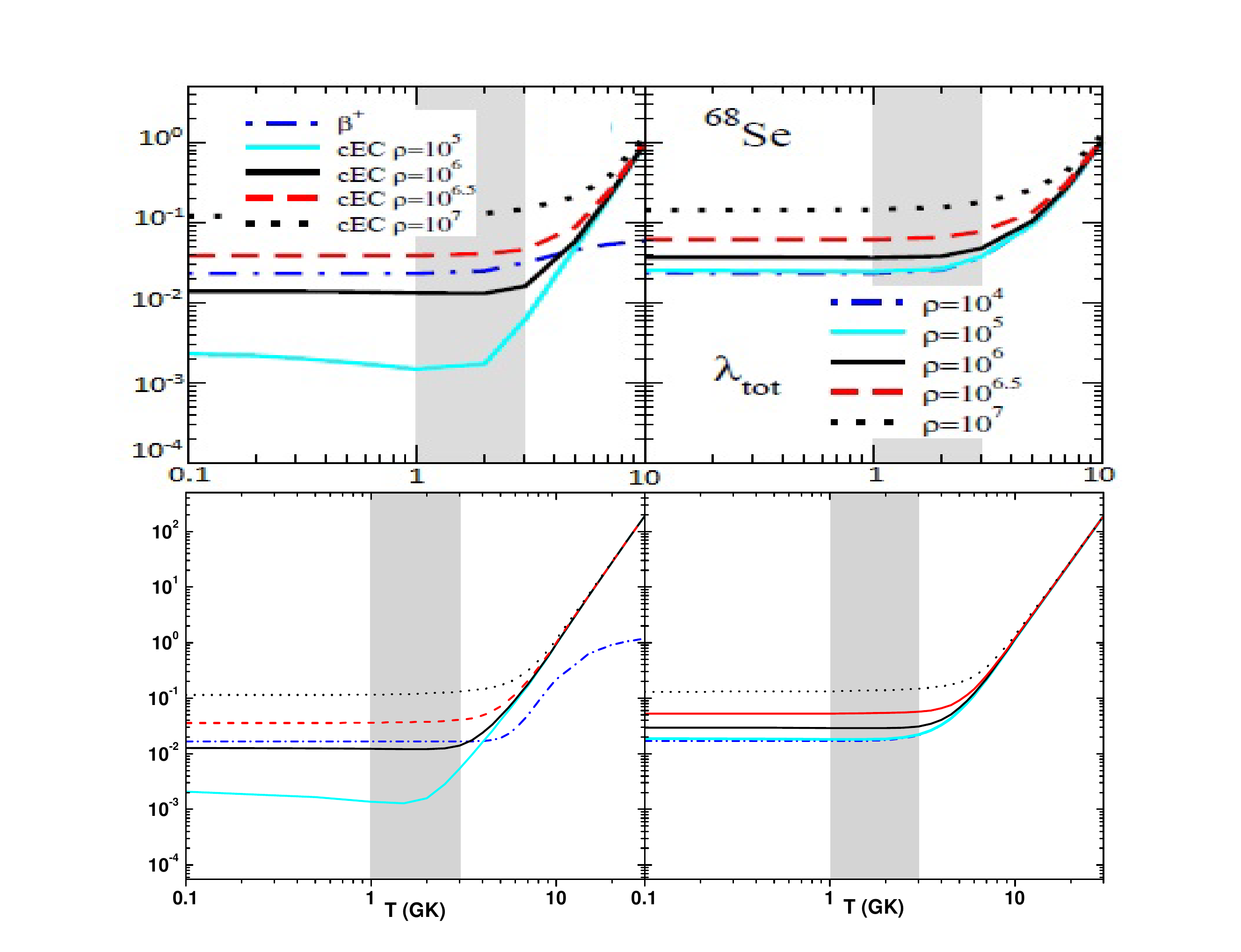}
\caption {(Color online) Same as Fig.~\ref{fig3} but for $^{68}$Se.}
\label{fig5}
\end{center}
\end{figure}

\begin{figure}
\begin{center}
\includegraphics[width=4.3in,height=4.3in]{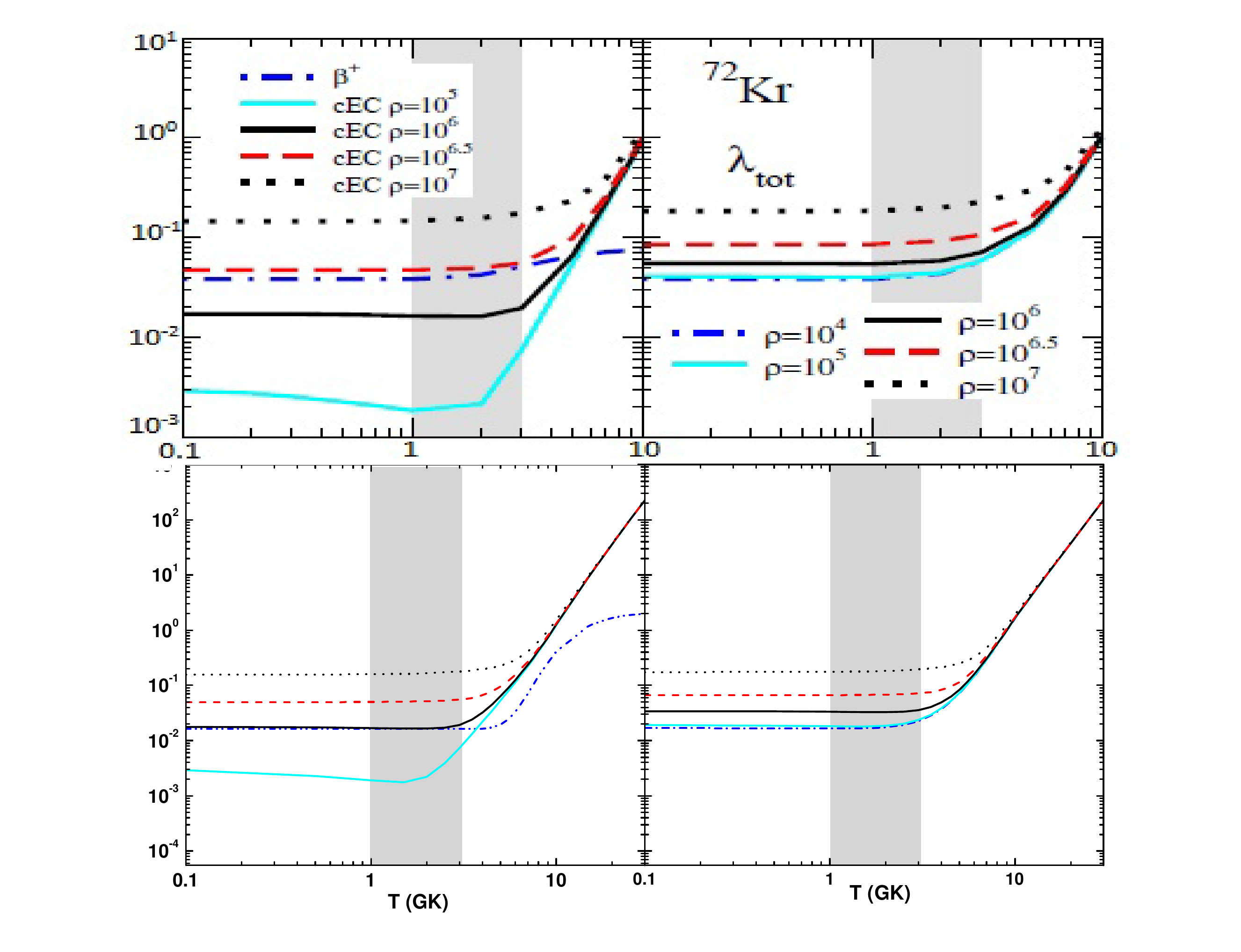}
\caption {(Color online) Same as Fig.~\ref{fig3} but for $^{72}$Kr.}
\label{fig6}
\end{center}
\end{figure}

\begin{figure}
\begin{center}
\includegraphics[width=4.3in,height=4.3in]{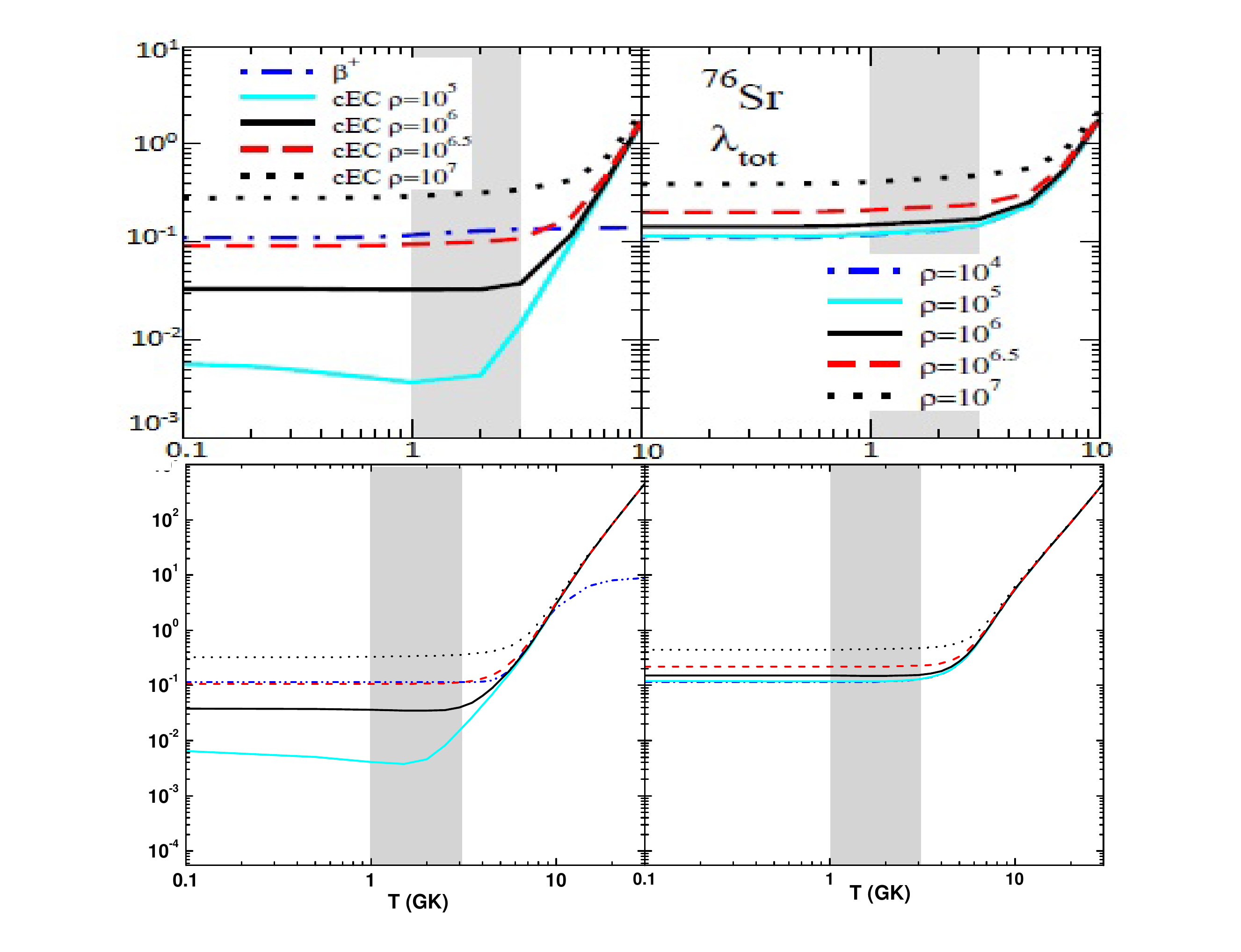}
\caption {(Color online) Same as Fig.~\ref{fig3} but for $^{76}$Sr.}
\label{fig7}
\end{center}
\end{figure}

The current \mbox{pn-QRPA} model was used earlier to calculate
$rp$-process weak-interaction mediated rates of waiting-point nuclei
\cite{Nab12} but using different model parameters discussed earlier.
Figs.~\ref{fig3} to~\ref{fig7} show the calculated weak-interaction
mediated rates for the WP nuclei as a function of the stellar
temperature and the density. The left-panels of these figures show
the stellar electron capture (cEC) and positron decay ($\beta^{+}$)
rates as a function of the stellar temperature for selected density
of 10$^{5}$, 10$^{6}$, 10$^{6.5}$ and 10$^{7}$ g.cm$^{-3}$
(pertinent to $rp$-process conditions). The positron decay rates
remain constant as the stellar density increases by two orders of
the magnitude. The right panels show the total sum of these two
rates. The upper panels are the Skyrme HF+BCS+QRPA calculation of
Sarriguren and reproduced from \cite{Sar11} whereas the lower panels
depict our results. Sarriguren calculated his rates only up to
stellar temperature of 10 GK. We performed our calculation up to 30
GK.  The shaded region is the temperature range considered to be
most relevant with the $rp$-process \cite{Sar11}. Fig.~\ref{fig3}
shows the comparison of the stellar rates for $^{60}$Zn. It can be
noted that our positron decay rates are around a factor 3 less than
Sarriguren rates under $rp$-process conditions. At temperatures of
10 GK our positron decay rates are factor 6 bigger. Our calculated
electron capture rates are in reasonable comparison with Sarriguren
calculation for $rp$-process conditions. Our calculated total rates
are factor 2 smaller for $rp$-process temperatures at density
(10$^{4}$ - 10$^{5}$) g.cm$^{-3}$ and in very good agreement with
Sarriguren calculated total rates at density(10$^{6}$ - 10$^{7}$)
g.cm$^{-3}$. Another feature to be noted is that for $rp$-process
conditions, our calculated positron decay rates are factor (1~--~6)
bigger at $\rho$ = 10$^{5}$ gm.cm$^{-3}$ when compared with our
calculated electron capture rates. However as stellar density
reaches $\rho$ = 10$^{7}$ gm.cm$^{-3}$, the calculated electron
capture rates increases and are factor (16 -- 19) bigger than the
competing positron decay rates.

Fig.~\ref{fig4} shows comparison of the stellar weak rates for
$^{64}$Ge. Here we note that our electron capture and positron decay
rates are in good agreement with Sarriguren rates for $rp$-process
conditions. At high temperature of 10 GK our electron capture
(positron decay) rates are factor 2 (5) bigger. Our calculated
positron capture rates are up to factor 20 (2) bigger than the
electron capture rates at $\rho$ = 10$^{5}$ (10$^{6}$) gm.cm$^{-3}$
for $rp$-process temperature range and are factor 4 smaller  at
$\rho$ = 10$^{7}$ gm.cm$^{-3}$.

For the case of $^{68}$Se, Fig.~\ref{fig5} shows that our rates are
up to three times bigger the corresponding Sarriguren rates for
$rp$-process conditions. At soaring temperatures around 10 GK our
positron decay rates are up to an order of the magnitude bigger
whereas the calculated electron capture rates are factor 4 bigger.
Once again our calculated electron capture rates are factor (7 -- 8)
bigger than the competing \mbox{pn-QRPA} positron decay rates at
$\rho$ = 10$^{7}$ gm.cm$^{-3}$ and $T$ = 1 -- 3 GK.

Comparison of the electron capture rates under $rp$-process
conditions for the case of $^{72}$Kr is almost perfect (see
Fig.~\ref{fig6}). Our calculated positron decay rates are half those
calculated by Sarriguren under similar physical conditions. At high
temperatures our positron decay rates are factor 5 bigger. For
$rp$-process conditions, the \mbox{pn-QRPA} calculated electron
capture rates are same as the $\beta^{+}$ decay rates at $\rho$ =
10$^{6}$ gm.cm$^{-3}$. At higher stellar density of $\rho$ =
10$^{7}$ gm.cm$^{-3}$, the calculated electron capture rates are an
order of the magnitude bigger.

Fig.~\ref{fig7} finally shows that for the prolate nucleus
$^{76}$Sr, Sarriguren calculated rates are in perfect agreement with
our calculation under $rp$-process conditions. At high temperatures
our positron decay (electron capture) rates are factor 18 (2)
bigger. For $rp$-process temperatures, our calculated positron decay
rates is factor 3 bigger at a stellar density of 10$^{6}$
gm.cm$^{-3}$ compared to our electron capture rates. When the
density increases by an order of the magnitude our electron capture
rates are three times the calculated $\beta^{+}$ decay rates.

Figs.~\ref{fig3} to~\ref{fig7} show that our calculated rates are
enhanced at high temperatures as compared to the Skyrme HF+BCS+QRPA
calculation. At high temperatures the calculated electron capture
rates are more than an order of magnitude bigger in comparison with
the competing positron decay rates. Convergence of the rate
calculation is in order as temperature increases to 10 GK and beyond
(see Eq. (\ref{tot})) due to finite occupation probability of parent
excited states. We took 200 initial and 300 final states in our rate
calculation which guaranteed satisfactory convergence in our rate
calculation. We note that due to the availability of a huge model
space (up to 7 major oscillator shells) in our \mbox{pn-QRPA} model,
convergence was achieved in our rate calculations for excitation
energies well in excess of 10 MeV. On the other hand the
self-consistent approach of the Skyrme HF+BCS+QRPA calculation
forces one to use limited configuration spaces which might lead to
convergence problem in rate calculation. Table~\ref{ta2} shows the
excited states contribution to the total electron capture and
positron decay rates under stellar conditions. It is noted that as
stellar temperature soars to 30 GK, a sizeable contribution comes
from the excited states. The excited states contribution is around
two orders of magnitude bigger for positron decay rates when
compared with the corresponding stellar electron capture rates. The
table also shows that under $rp$-process conditions the
contributions from excited states are almost negligible.

\section{Conclusions}
\label{sec:level7}

Accurate estimate of the positron decay and electron capture rates
of the WP neutron-deficient medium mass nuclei are required for a
better understanding of the $rp$-process. Incidently nuclear
deformation is believed to play a crucial role in determining the
strength distributions of the $\beta$-decay for these medium mass
nuclei. Besides there exist a wide variety of the nuclear shapes
displayed in this region. In this project we selected five,
$N$~=~$Z$, the WP nuclei, namely $^{60}$Zn, $^{64}$Ge, $^{68}$Se,
$^{72}$Kr and $^{76}$Sr, to study their nuclear structure properties and geometric shapes and to
calculate their half-lives and associated stellar weak rates. The
nuclear shape was determined using the \mbox{PES} of the
\mbox{IBM-1} model. The model calculated essentially spherical
nuclei for the first four cases and a $\beta_{min}$ value of 0.86
for $^{76}$Sr. As we were not able to find in literature an exact
formula to convert our IBM calculated $\beta$ to deformations for
use in Nilsson model, we decided to use the deformations calculated
from the relativistic mean-field theory.  We then performed a
\mbox{pn-QRPA} calculation in a huge model space of 7$\hbar\omega$
to calculate the half-lives of these neutron-deficient WP nuclei.
The calculated half-lives were in very good agreement with the
experimental half-lives determined from the recent atomic mass
evaluation AME2012.

We presented the BGT strength distribution for the WP nuclei and
compared with previous calculation and measurements wherever
possible. Our calculated strength was in decent comparison with the
measured data. For the case of $^{60}$Zn, the Skyrme HF+BCS+QRPA
calculation of total GT strength, performed by Sarriguren, was
slightly bigger. For remaining cases our calculated total GT
strength was bigger. Our calculated total strength was in good
comparison with Sarriguren calculation for the case of $^{60}$Zn and
$^{76}$Sr.  We then calculated the positron decay and electron
capture rates for these proton rich WP nuclei in stellar matter with
special focus on $rp$-process conditions. Our stellar rates were
also compared with those performed by Sarriguren and agreed to
within a factor of two for $rp$-process conditions.

Our calculation showed that electron capture rates compete well with
the positron decay rates under $rp$-process conditions. In essence
the electron capture rates were bigger, by more than an order of the
magnitude, than the competing $\beta^{+}$ decay rates for
temperature of 3 GK and density $\rho$ = 10$^{7}$ gm.cm$^{-3}$. Our
findings reiterates the fact that electron capture rates on WP
proton-rich nuclei form an integral part of the weak rates under
$rp$-process conditions and must not be neglected in nuclear network
calculations.

\section*{Acknowledgements}
J.-U. Nabi wishes to acknowledge the support provided by the Higher
Education Commission (Pakistan) through the HEC Project No. 20-3099.



\bibliographystyle{elsarticle-num}



\newpage


\begin{table}
\centering
\begin{tabular}{cccccc}
\hline
~~~~Nucelus~~~~&~~~~$^{60}$Zn~~~~&~~~~$^{64}$Ge~~~~&~~~~$^{68}$Se~~~~&~~~~$^{72}$Kr~~~~&~~~~$^{76}$Sr\\
\hline
$~~~~N~~~~$&~~~~2~~~~&~~~~4~~~~&~~~~6~~~~&~~~~8~~~~&~~~~10\\
$~~~~\beta$[IBM]~~~~&~~~~0~~~~&~~~~0~~~~&~~~~0~~~~&~~~~0~~~~&~~~~0.86\\
$\beta$ \cite{Lal99} & +0.170 & +0.217 & -0.285 & -0.358 & +0.410\\
\hline
\end{tabular}
\caption{Calculated values of nuclear deformation using the IBM-1
Model and those using the relativistic mean-field theory
\cite{Lal99}. $N$ is boson number.} \label{ta1ii}
\end{table}

\begin{table}
\centering
\begin{tabular}{cccccc}
\hline
 & N &$\epsilon ^{[1]}$&$a_{2}^{[1]}$&$\overline{\chi}^{[2]}$&$\sigma ^{[3]}$\\
\hline
$^{60}$Zn& 2 & 883 & -52.3 & -0.35 & 0 \\
$^{64}$Ge& 4 & 914.1 & -77.2 & -0.35 & 15 \\
$^{68}$Se& 6 & 1116.8 & -59.8 & -0.35 & 12 \\
$^{72}$Kr& 8 & 1026.85 & -26.4 & -1.3 & 60 \\
$^{76}$Sr& 10 & 300.6 & -71.1 & -0.25 & 4 \\
\hline
\end{tabular}
\caption{The fitted parameters of the IBM-1 Hamiltonian~( Eq.
\ref{e_ham}). $[1]$ given in units of keV, $[2]$ dimensionless,
$[3]$ the root-mean-square (\emph{rms}) deviation in units of keV.}
\label{ta1}
\end{table}

\begin{table}
\centering
\begin{tabular}{ccccccccccc}
\hline
$~\overline{\chi}~~$&~~-0.2~~&~~-0.4~~&~~-0.6~~&~~-0.7~~&~~-0.8~~&~~-0.9~~&~~-1.0~~&~~-1.1~~&~~-1.2~~&~~-1.~\\
\hline
$~\sigma~~$&~~69~~&~~68~~&~~67~~&~~66~~&~~65~~&~~64~~&~~63~~&~~62~~&~~61~~&~~60~\\
\hline
\end{tabular}
\caption{Detail fitted procedure for $\overline{\chi}$ value of
$^{72}$Kr nucleus.} \label{ta1i}
\end{table}

\begin{table}
\centering
\begin{tabular}{@{}lllll}
\hline
~~~~~~~~&~~~~~~~~Exp.~~~~~~~~&~~~~~~~~IBM-1~~~~~~~~~~~~~~~~\\
\hline
~~~~~~~~$^{60}$Zn~~~~~~~~&~~~~~~~~ - ~~~~~~~~& ~~~~~~~~ 2.04~~~~~~~~  \\
~~~~~~~~$^{64}$Ge~~~~~~~~&~~~~~~~~ - ~~~~~~~~& ~~~~~~~~ 5.26~~~~~~~~ \\
~~~~~~~~$^{68}$Se~~~~~~~~&~~~~~~~~ 4.45 $(0.66)$ ~~~~~~~~& ~~~~~~~~ 8.94~~~~~~~~ \\
~~~~~~~~$^{72}$Kr~~~~~~~~&~~~~~~~~ 10.14 $(1.42)$ ~~~~~~~~& ~~~~~~~~ 12.41~~~~~~~~ \\
~~~~~~~~$^{76}$Sr~~~~~~~~&~~~~~~~~ - ~~~~~~~~& ~~~~~~~~ 25.68~~~~~~~~ \\
\hline
\end{tabular}
\caption{Experimental and calculated B(E2: $2^+_1\rightarrow0^+_1$)
values in units of $10^{-2}$ $e^{2}b^{2}$}\label{ta1iii}
\end{table}

\begin{table}
\centering \normalsize\begin{tabular}{c|cccc}

 & &\emph{$\mathbf{^{60}Zn}$} & & \\
$\mathbf{(\rho ,T_{9})}$& $\mathbf{\lambda_{ec}(G)}$ &
$\mathbf{R_{ec}(G/T)}$ &$\mathbf{\lambda_{pd}(G)}$ &
$\mathbf{R_{pd}(G/T)}$ \\\hline

($10^{6.5}$,1)   & 2.32E-02 & 1.00E+00 &    4.87E-03 & 1.00E+00 \\
($10^{6.5}$,1.5) & 2.39E-02 & 1.00E+00 &    4.87E-03 & 1.00E+00 \\
($10^{6.5}$,2)   & 2.49E-02 & 1.00E+00 &    4.87E-03 & 1.00E+00 \\
($10^{6.5}$,2.5) & 2.62E-02 & 1.00E+00 &    4.87E-03 & 1.00E+00 \\
($10^{6.5}$,3)   & 2.80E-02 & 1.00E+00 &    4.87E-03 & 9.98E-01 \\
($10^{6.5}$,30)  & 7.65E+00 & 4.65E-02 &    2.11E-04 & 2.29E-04 \\
\hline & &\emph{$\mathbf{^{64}Ge}$} & & \\

($10^{6.5}$,1)   & 1.08E-02 & 1.00E+00 &    7.81E-03 & 1.00E+00 \\
($10^{6.5}$,1.5) & 1.10E-02 & 1.00E+00 &    7.81E-03 & 1.00E+00 \\
($10^{6.5}$,2)   & 1.12E-02 & 1.00E+00 &    7.81E-03 & 1.00E+00 \\
($10^{6.5}$,2.5) & 1.14E-02 & 1.00E+00 &    7.81E-03 & 1.00E+00 \\
($10^{6.5}$,3)   & 1.15E-02 & 1.00E+00 &    7.81E-03 & 9.96E-01 \\
($10^{6.5}$,30)  & 6.69E+00 & 3.52E-02 &    2.96E-04 & 5.44E-04 \\
\hline & &\emph{$\mathbf{^{68}Se}$} & & \\

($10^{6.5}$,1)   & 3.63E-02 & 1.00E+00 &    1.68E-02 & 1.00E+00 \\
($10^{6.5}$,1.5) & 3.68E-02 & 1.00E+00 &    1.68E-02 & 1.00E+00 \\
($10^{6.5}$,2)   & 3.75E-02 & 1.00E+00 &    1.68E-02 & 1.00E+00 \\
($10^{6.5}$,2.5) & 3.87E-02 & 1.00E+00 &    1.68E-02 & 1.00E+00 \\
($10^{6.5}$,3)   & 4.05E-02 & 1.00E+00 &    1.68E-02 & 9.99E-01 \\
($10^{6.5}$,30)  & 9.35E+00 & 4.95E-02 &    7.70E-04 & 6.56E-04 \\
\hline & &\emph{$\mathbf{^{72}Kr}$} & & \\

($10^{6.5}$,1)   & 4.99E-02 & 1.00E+00 &    1.63E-02 & 1.00E+00 \\
($10^{6.5}$,1.5) & 5.05E-02 & 1.00E+00 &    1.63E-02 & 1.00E+00 \\
($10^{6.5}$,2)   & 5.14E-02 & 1.00E+00 &    1.63E-02 & 1.00E+00 \\
($10^{6.5}$,2.5) & 5.28E-02 & 1.00E+00 &    1.63E-02 & 1.00E+00 \\
($10^{6.5}$,3)   & 5.52E-02 & 1.00E+00 &    1.63E-02 & 9.99E-01 \\
($10^{6.5}$,30)  & 9.18E+00 & 4.08E-02 &    6.94E-04 & 3.51E-04 \\
\hline & &\emph{$\mathbf{^{76}Sr}$} & & \\

($10^{6.5}$,1)   & 1.05E-01 & 1.00E+00 &    1.14E-01 & 1.00E+00 \\
($10^{6.5}$,1.5) & 1.06E-01 & 1.00E+00 &    1.14E-01 & 1.00E+00 \\
($10^{6.5}$,2)   & 1.07E-01 & 1.00E+00 &    1.14E-01 & 1.00E+00 \\
($10^{6.5}$,2.5) & 1.09E-01 & 1.00E+00 &    1.14E-01 & 1.00E+00 \\
($10^{6.5}$,3)   & 1.13E-01 & 1.00E+00 &    1.14E-01 & 9.99E-01 \\
($10^{6.5}$,30)  & 1.18E+01 & 2.64E-02 &    5.02E-03 & 5.75E-04 \\
\end{tabular}
\caption{The \textit{ground state} electron and positron decay
rates, $\lambda_{ec}(G)$, $\lambda_{pd}(G)$, respectively, for
$^{60}$Zn, $^{64}$Ge, $^{68}$Se, $^{72}$Kr and $^{76}$Sr in units of
$sec^{-1}$. Given also are the ratios of the ground state capture
and decay rates to total rate, $R_{ec}(G/T)$, $R_{pd}(G/T)$,
respectively. The first column gives the corresponding values of
stellar density, $\rho $ ($gcm^{-3}$), and temperature, $T_{9}$ (in
units of $10^{9}$ K), respectively.} \label{ta2}
\end{table}

\newpage

\clearpage
\clearpage
\clearpage
\clearpage
\clearpage
\clearpage
\clearpage
\clearpage
\clearpage
\clearpage
\clearpage
\clearpage

\end{document}